\documentclass[intlimits,twoside,a4paper]{article}

\usepackage{amsmath}
\usepackage{amssymb}
\usepackage{graphicx}
\usepackage[T2A]{fontenc}
\usepackage[cp1251]{inputenc}

\usepackage{color}
\usepackage{cmpj2}

\issue{2015}{18}{2}{23601}
\doinumber{10.5488/CMP.18.23601}

\title{Ab initio study of structural, electronic, and thermal properties of Ir$_{1-x}$Rh$_{x}$  alloys}%

\author[S. Ahmed \textsl{et al.}]{S. Ahmed\thanks{E-mail: shabir\_sehr@hotmail.com}, M. Zafar, M. Shakil, M.A. Choudhary} %
\address{
Simulation Laboratory, Department of Physics, The Islamia University of Bahawalpur, Bahawalpur 63100, Pakistan
}%

\date{Received August 4, 2014, in final form October 31, 2014}
\authorcopyright{Sh. Ahmed, M. Zafar, M. Shakil, M.A. Choudhary, 2015}

\begin{document}

\maketitle

\begin{abstract} \label{abstract}
The structural, electronic, mechanical and thermal  properties of Ir$_{1-x}$Rh$_{x}$ alloys were studied
systematically using ab initio density functional theory at different concentrations ($x = 0.00$, 0.25, 0.50, 0.75, 1.00).
A Special Quasirandom Structure method was used to make alloys having FCC structure with four atoms per unit cell.
The ground state properties such as lattice constant and bulk modulus were calculated to find the equilibrium atomic
position for stable alloys. The calculated ground state properties are in good agreement with the experimental and
previously presented other theoretical data. The electronic band structure and density of states were calculated
to study the electronic properties for these alloys at different concentrations. The electronic properties substantiate
the metallic behavior of alloys. The first principle density functional perturbation theory as implemented in
quasiharmonic approximation was used for the calculation of thermal properties. We have calculated the thermal
properties such as Debye temperatures, vibration energy, entropy, constant-volume specific heat and internal energy.
The ab initio linear-response method was used to calculate phonon densities of states.
\pacs 61.50f, 71.15Mb, 71.20Be
\end{abstract}

\section{Introduction}
Platinum group metals are promising candidates for a wide range of applications. Iridium (Ir), Rho\-dium (Rh) belong to this group and are the rarest elements found in the earth's crust. Their  incredible catalysis activity increases research interest in these rare elements \cite{1}. Iridium has high density, high melting temperature and the highest corrosion resistance. Its thermal stability at high temperatures and the highest shear modulus at room temperature  made it a superior metal \cite{2,3,4}. It has been investigated for carbon materials protective coatings \cite{5}, heavy-metal-ion sensors \cite{6} and Re-rocket thrusters \cite{7} applications. It is broadly used in organic chemistry for hydrogenation, hydroformulation, hydroboration, hydrosilylation, cycloadditions \cite{8,9}  and oxidation reactions for catalytic converters in automobiles. It remains a metal of high interest due to its unique properties for the scientific community \cite{10,11} during the past decades. Binary intermetallic alloys that contain a transition metal display interesting electronic, structural, optical and thermal properties. Recently, Ir- and Rh-base alloys are known as ultra-high temperature materials due to their high-melting temperatures, good high-temperature strengths and good oxidation resistances \cite{12,13,14}. Rh-base alloys have better oxidation resistance, lower density, lower thermal expansion coefficient and higher thermal conductivity than Ir-base alloys. These properties make Rh-base alloys more promising for ultra-high temperature gas turbine applications. To our knowledge, theoretical and experimental calculation are performed for Ir, Rh and their alloys with other metals but neither theoretical nor experimental calculations are available for Ir-Rh alloys. These alloys provide a class of systems exhibiting unique mechanical properties that make them attractive for structural applications \cite{15}.
In recent years, Ir-base super-alloys with good high-temperature properties have been developed as promising candidates for ultrahigh-temperature applications to replace the traditional Ni-base super-alloys, which have limited temperature capabilities due to the rather low melting point of Ni \cite{16,17,18}. These alloys are promising materials for high temperature and pressure applications, and currently they are being examined for use in diesel engine turbocharger rotors, high-temperature die and molds, hydroturbines, and cutting tools \cite{19}.
Due to the improvements in simulation techniques in material science, it is now possible to study materials from bulk to nano scale devices \cite{20}. The use of first principles calculations offers one of the most powerful tools for carrying out theoretical studies of a number of important physical and chemical properties of condensed matter with great accuracy \cite{21,22}. Electronic structure simulations based on density-functional theory (DFT) \cite{23,24} have been instrumental to this revolution, and their application has now spread outside a restricted core of researchers in condensed-matter theory and quantum chemistry, involving a vast community of end users with very diverse scientific backgrounds and research interests.  The knowledge of thermal properties is essential for the study of mechanical properties \cite{25}. To the best of our knowledge, there are no experimental and other theoretical data for comparison, so we consider the present results as a predictive study for the first time, which still awaits an experimental confirmation.

\section{Computational method}

The first principles investigation was performed using a pseudo-potential plane wave (PP-PW) method as implemented in QUANTUMES ESPRESSO \cite{26}. Special Quasirandom Structure method proposed by Zunger and coworkers \cite{32} was used to make FCC structure with four atoms per unit cell alloys. SQSs are specially designed for a small supercell that is computationally feasible for DFT calculations. Vanderbilt ultra soft pseudo-potential parameterized by Perdew and Zunger \cite{27,28} was used for calculations. We used ultrasoft Vanderbilt formalism with local density approximation (LDA) \cite{29} for the exchange correlation energy of electrons. The special $k$-points were integrated to the sampled Brillouin zone using the Monkhorst-Pack method \cite{31}. A special $k$-points mesh of $14\times14\times14$ was used to produce an irreducible Brillouin zone. Pseudo-wave functions were expanded in a plane wave basis set using the cut-off energy of 25~Ry for all concentrations. The chosen cut-off energy and $k$-point mesh ensure convergence with an accuracy of 10$^{-6}$~Ry. Density Functional Perturbation Theory (DFPT) and Quasi Harmonic Approximation (QHA) code developed by Baroni et al. \cite{26} were used to investigate the phonon density of states and thermal properties from optimized structure. The dynamical matrices at arbitrary wave vectors were obtained using the Fourier transformation-based interpolations to calculate thermal properties. A $12\times12\times12$ $q$ point mesh was implemented to obtain dynamical matrices of force constants in the irreducible Brillouin zone at $\Gamma$ point. The linear response approach method was used to obtain the curves of phonon density of states within the framework of DFPT \cite{26} as implemented in Quantum ESPRESSO.
The parameters such as cell dimension, energy cut-offs and the $k$-points used in the study were obtained from their respective convergence tests. These convergence parameters provide a fast way toward an optimized structure.

\vspace{5mm}

\section{Results and discussion}

\subsection{Structural properties}

We have found the ground state properties of Ir$_{1-x}$Rh$_{x}$  alloys at $x= 0$, 0.25, 0.5, 0.75, and 1.00. The total energy per unit cell was computed at various lattice parameters to find the equilibrium lattice constant. The equilibrium lattice constant was found by minimizing the total energy of system. Murnaghan's equation of state was used to evaluate the optimized lattice constants and bulk moduli \cite{33}.

\begin{table}
\caption{Calculated lattice parameters and bulk modulus of Ir$_{1-x}$Rh$_{x}$ alloys
compared with experimental and other theoretical results at $x= 0$, 0.25, 0.5, 0.75, and 1.00.\label{tab1}}
\vspace{1ex}
\begin{center}
\begin{tabular}{|c|c|c|c|c|c|c|}
\hline
     Composition & \multicolumn{3}{|c|}{Lattice constant (Angstrom)} & \multicolumn{3}{|c|}{Bulk modulus (Gpa)}\\
\hline
$x$ & This work & Experimental & Other  & This work & Experimental & Other \\
& & & calculation & & & calculation\\
			\hline\hline
			0.00 & 3.831 & 3.84$^a$ & 3.88$^b$ & 360 & 355$^e$ & 386$^g$ \\
			0.25 & 3.81394 &  &  & 337.375 &  & \\
			0.50 & 3.80373 &  &  & 315 &  & \\
			0.75 & 3.79589 &  &  & 294.5 &  & \\
			1.00 & 3.792 & 3.803$^c$ & 3.847$^d$ & 269.5 & 268.7$^f$ & 259.6$^h$ \\
\hline
\end{tabular}
\end{center}
$^a$ reference \cite{34}, \qquad $^b$ reference \cite{35}, \qquad $^c$ reference \cite{36}, \qquad $^d$ reference \cite{37}, \\
$^e$ reference \cite{38}, \qquad $^f$ reference \cite{39}, \qquad $^h$ reference \cite{40}, \qquad $^g$ reference \cite{41}.
\end{table}

\begin{figure}
\includegraphics[width=0.48\textwidth]{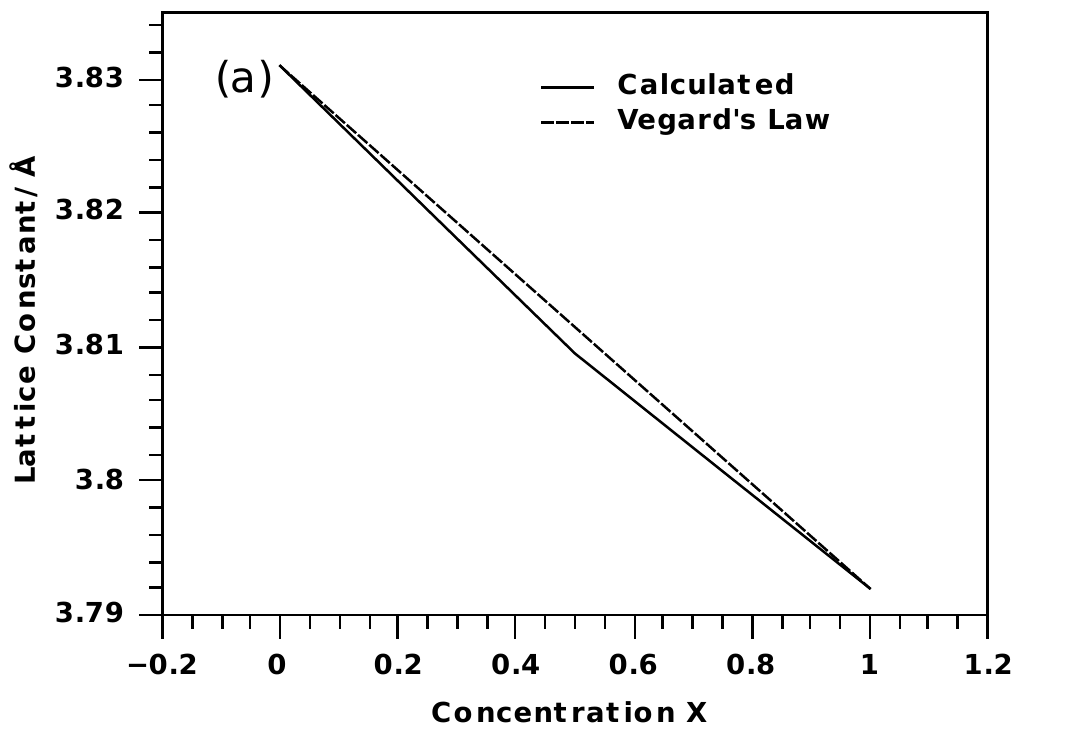}%
\hfill%
\includegraphics[width=0.48\textwidth]{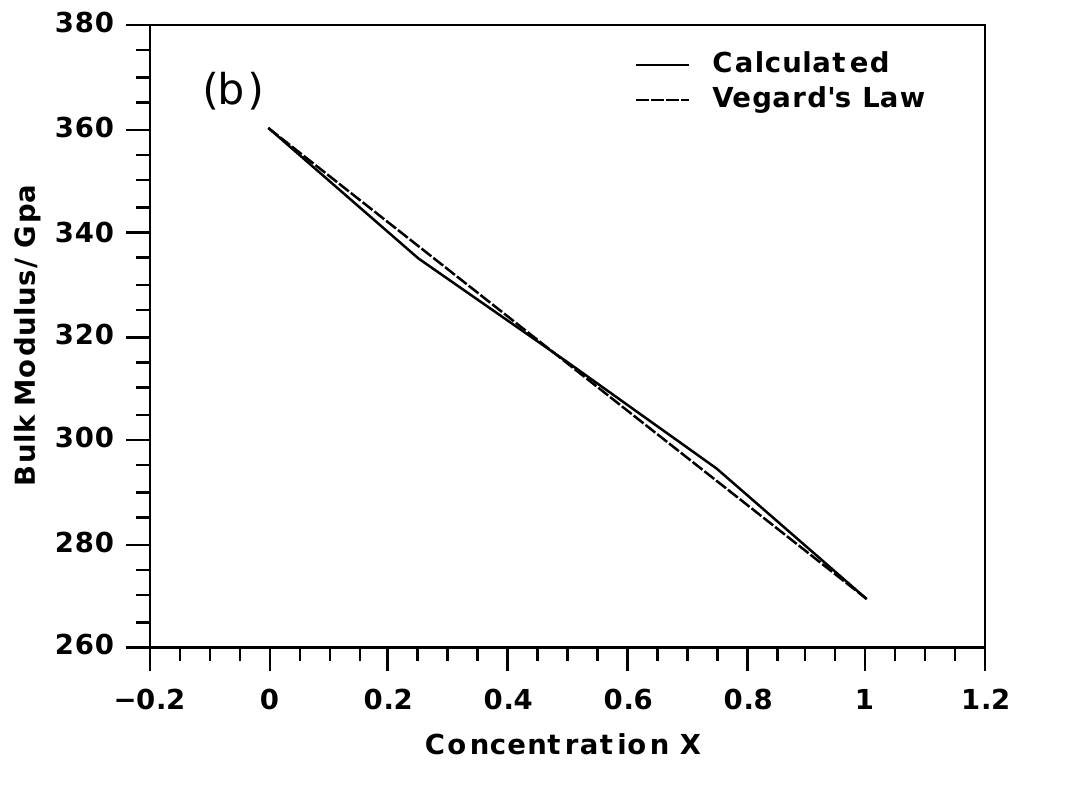}%
\caption{Calculated lattice parameters and bulk modulus in comparison with Vegard's law at experimental values for Ir$_{1-x}$Rh$_{x}$ alloys: (a) lattice constant, (b) bulk modulus.}
\label{fig1}
\end{figure}

\begin{figure}[!h]
\includegraphics[width=0.48\textwidth]{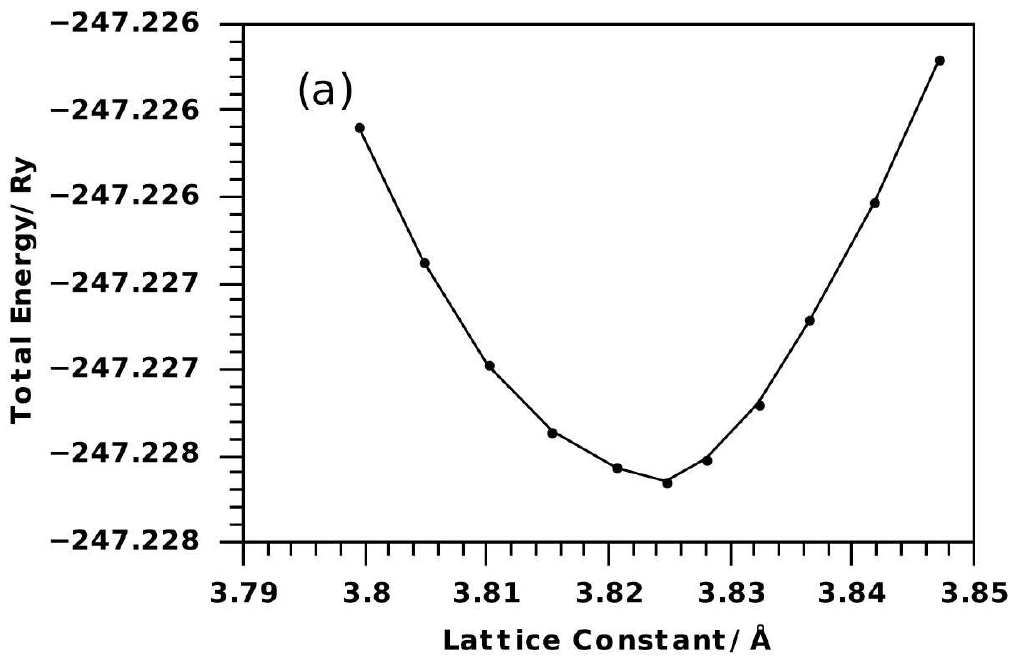}%
\hfill%
\includegraphics[width=0.48\textwidth]{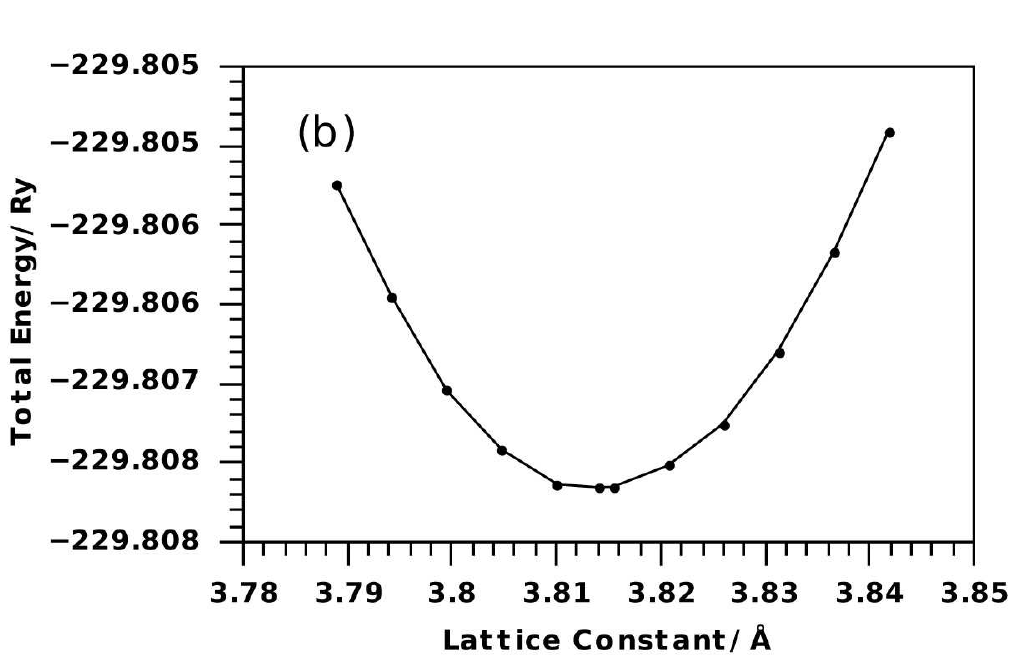}%
\\
\includegraphics[width=0.48\textwidth]{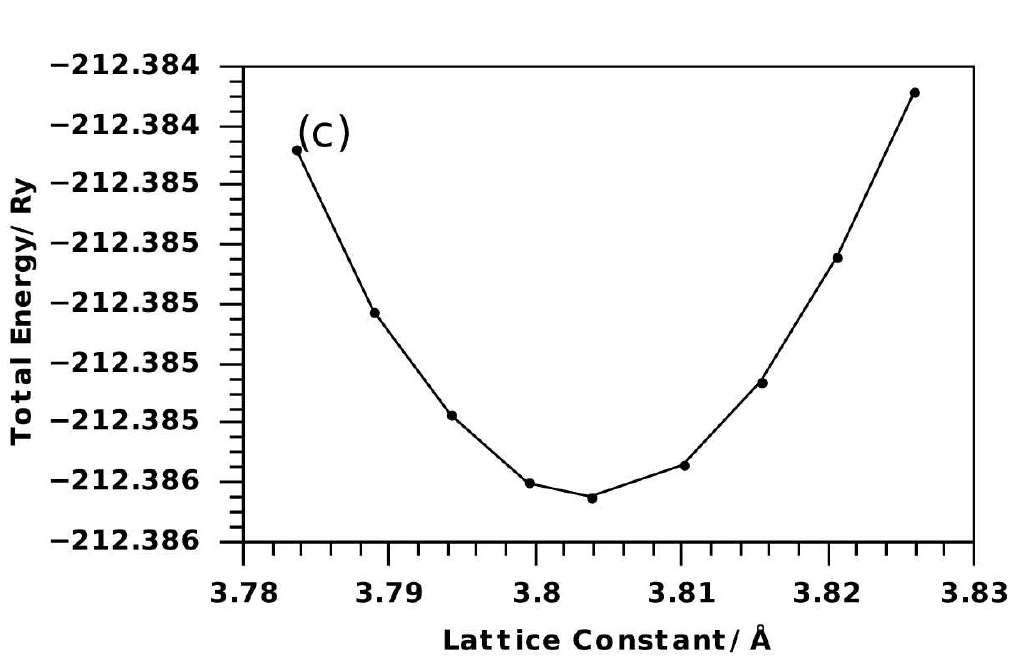}%
\hfill%
\includegraphics[width=0.48\textwidth]{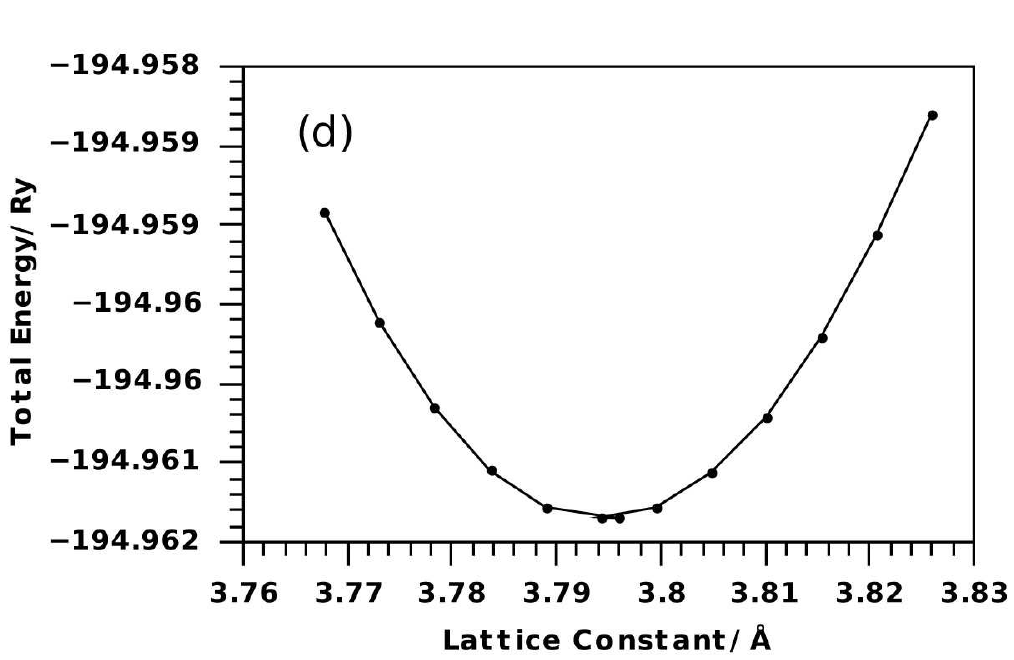}%
\\
\centerline{\includegraphics[width=0.48\textwidth]{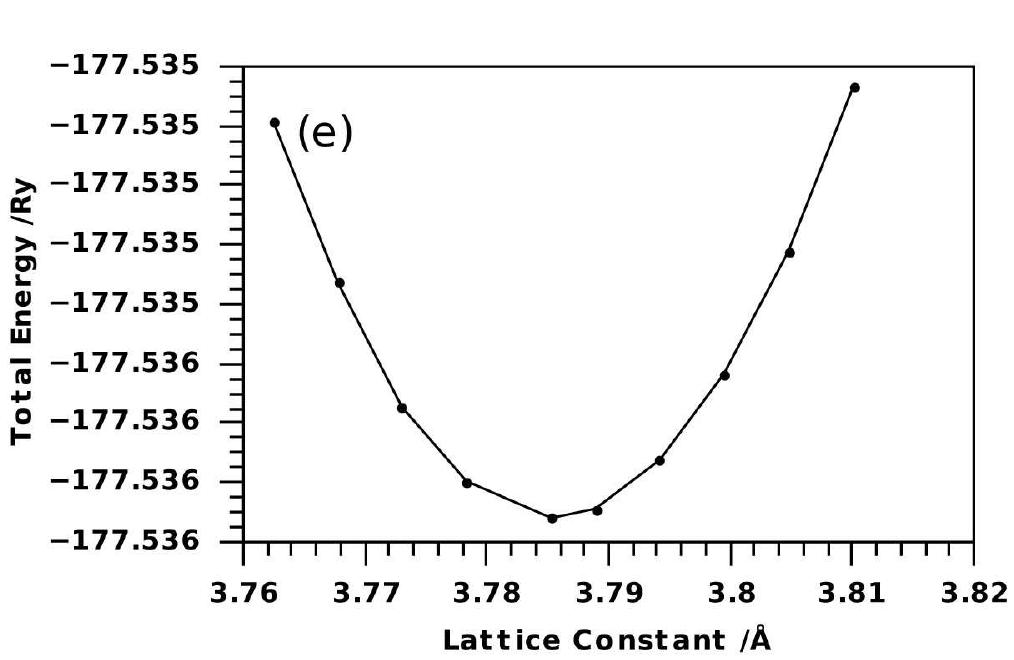}}
 \caption{Structural optimization plots for Ir$_{1-x}$Rh$_{x}$ alloys: (a) at $x= 0.00$, (b) at $x= 0.25$, (c) at $x= 0.50$, (d) at $x= 0.75$ and (e) at $x= 1.00$ concentrations.}
\label{fig2}
\end{figure}

The structural parameters i.e., lattice constants and bulk modulus for pure Ir, Rh and their alloys were calculated and reported in table~\ref{tab1} along with experimental and other theoretical results. The optimized lattice  curves  for Ir$_{1-x}$Rh$_{x}$ alloys at $x= 0$, 0.25, 0.5, 0.75, and 1.00 are shown in figure~\ref{fig2}. Figure~\ref{fig1} shows the variation of lattice constants and bulk modulus for Ir$_{1-x}$Rh$_{x}$ evaluated using Vegard's law. It is seen that with the change of Rh concentration in Ir$_{1-x}$Rh$_{x}$ alloys, the lattice constant and bulk modulus show a considerable change from Vegard's law which may be due to the lattice mismatch in these alloys.


The calculated results are compared with experimental and other theoretical results for pure Ir and Rh. However, the experimental or theoretical results for their alloys are not yet available in the literature for comparison. These calculations may be a prediction for researchers for future.  The values of the calculated lattice constant and the bulk modulus for Ir$_{1-x}$Rh$_{x}$ alloys turned out to decrease with an increase of  the Rh concentration.

\subsection{Electronic properties}

\begin{figure}[!h]
\includegraphics[width=0.48\textwidth]{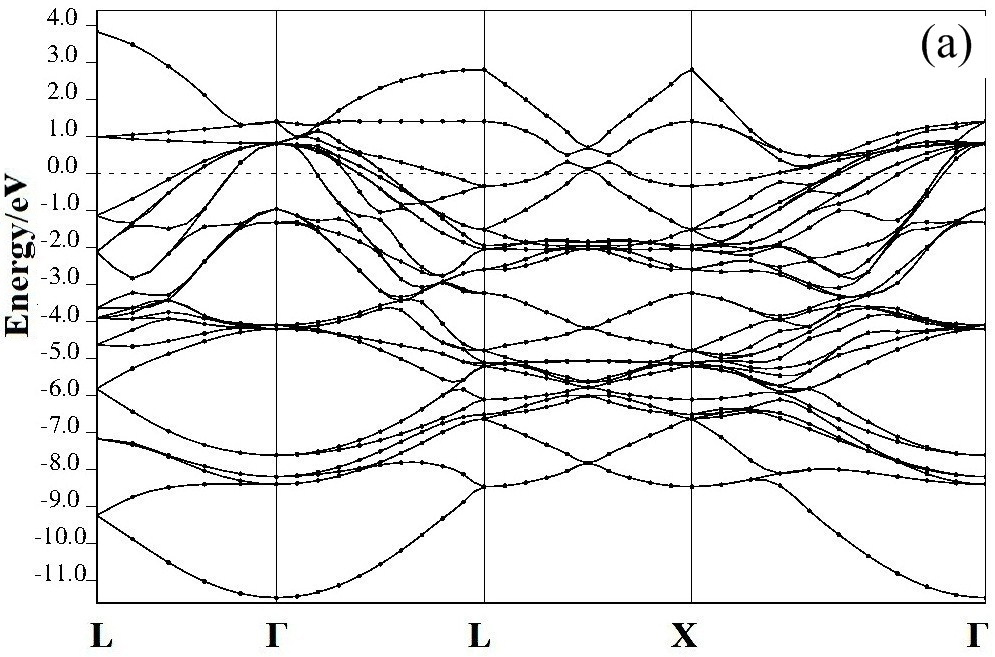}%
\hfill%
\includegraphics[width=0.48\textwidth]{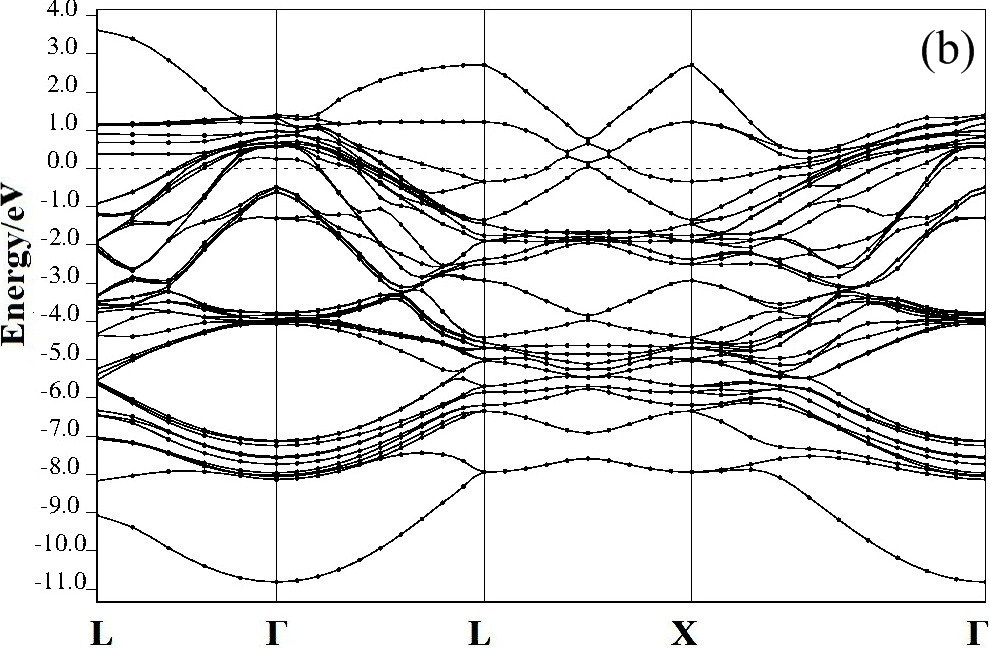}
\\
\includegraphics[width=0.48\textwidth]{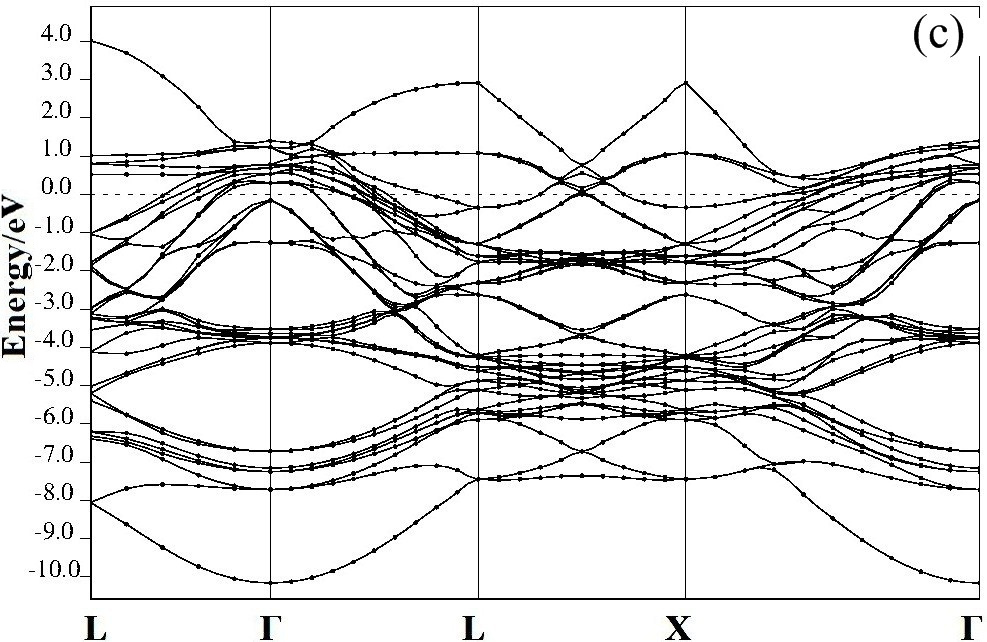}%
\hfill%
\includegraphics[width=0.48\textwidth]{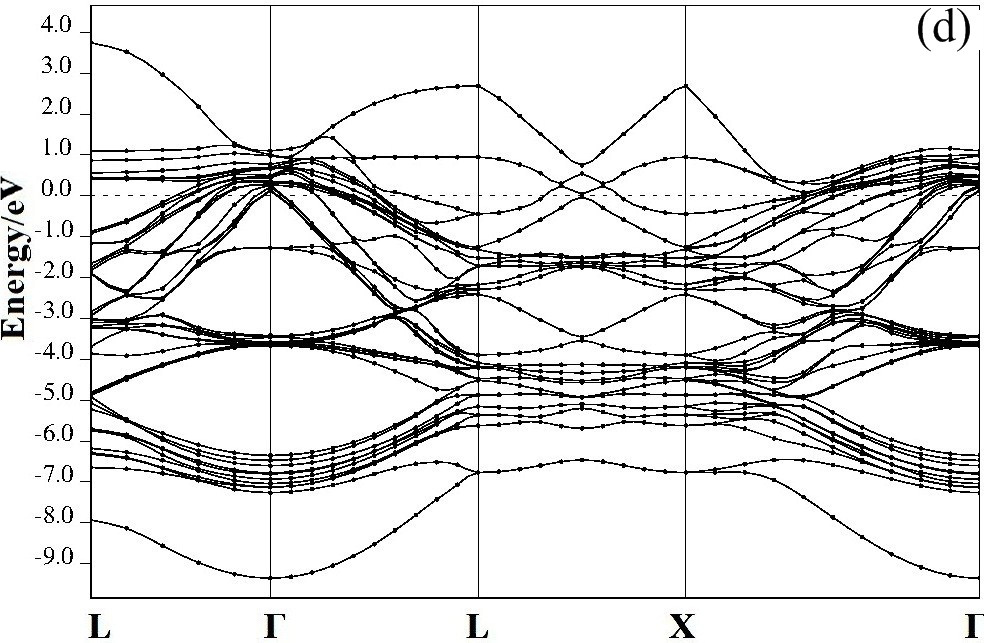}
\\
\centerline{\includegraphics[width=0.48\textwidth]{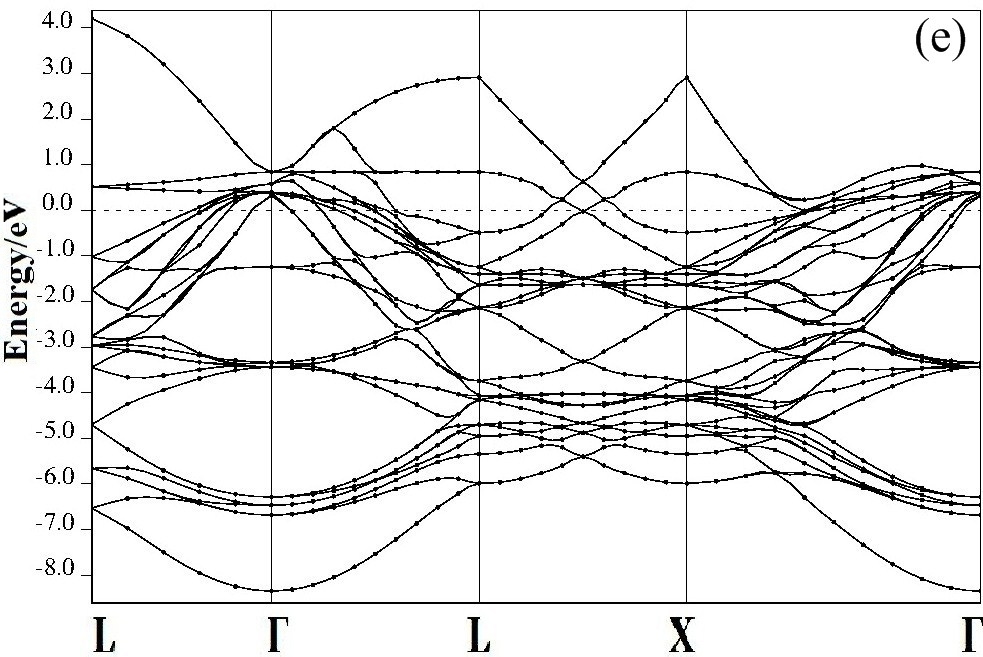}}
 \caption{Calculated band structures for Ir$_{1-x}$Rh$_{x}$ alloys: (a) at $x= 0.00$, (b) at $x= 0.25$, (c) at $x= 0.50$, (d) at $x= 0.75$ and (e) at $x= 1.00$ concentrations.}
\label{fig3}
\end{figure}

Electronic properties  depend on the electronic configuration in materials, particularly on the existence of prohibited regions of energy and on the magnitude in their electronic excitation spectra. Detailed band-structure calculations are needed to understand the electronic properties of any material. Thus, the band structures for Ir$_{1-x}$Rh$_{x}$ along various  high symmetry directions were calculated at equilibrium lattice constants. The electronic band structures are shown in figure~\ref{fig3}. The Fermi level for the band structure was  set to be 0~eV. It can be seen from figure~\ref{fig3}~(a)--(e) that electronic bands for all concentrations  overlap at the Fermi level. The electronic structures calculated for different concentrations  clearly show the metallic nature of the materials since the energy bands intersect at the Fermi level. Conductivity in Ir metal  is dominated by overlapping of $s$, $p$ and $d$ bands.

For Rh metal, the conductivity is dominated by a narrow $d$-band with some empty states overlapped by a broad free electron $s$-band. Their alloys show a metallic nature due to $s$, $p$, $d$ of Ir atoms and $s$, $d$ of Rh atoms bands overlapping. So, these alloys lead to archetypical transition metals, in which a narrow $d$-band containing some empty states was overlapped by a broad free-electron like $s$-band which dominated the conductivity \cite{42}.

The calculated electronic band structures along the principal symmetry directions are presented in figure~\ref{fig3}~(a)--(e). The overall band structure shows a similar metallic nature. The main hybridization remains between $d$ bands of both Ir and Rh when mixed in order to form their alloys. The Fermi energy decreases with an increase of  Rh percentage in Ir$_{1-x}$Rh$_{x}$ alloys. The calculated  electronic band structure is in good agreement with the previous result for pure Ir and Rh \cite{43,44}. From the calculated electronic band structure for Ir$_{1-x}$Rh$_{x}$ alloys, it can be seen that the band overlapping increases with Rh concentration  which causes a decrease of resistivity with Rh concentration \cite{45}.

\begin{figure}[!h]
\includegraphics[width=0.48\textwidth]{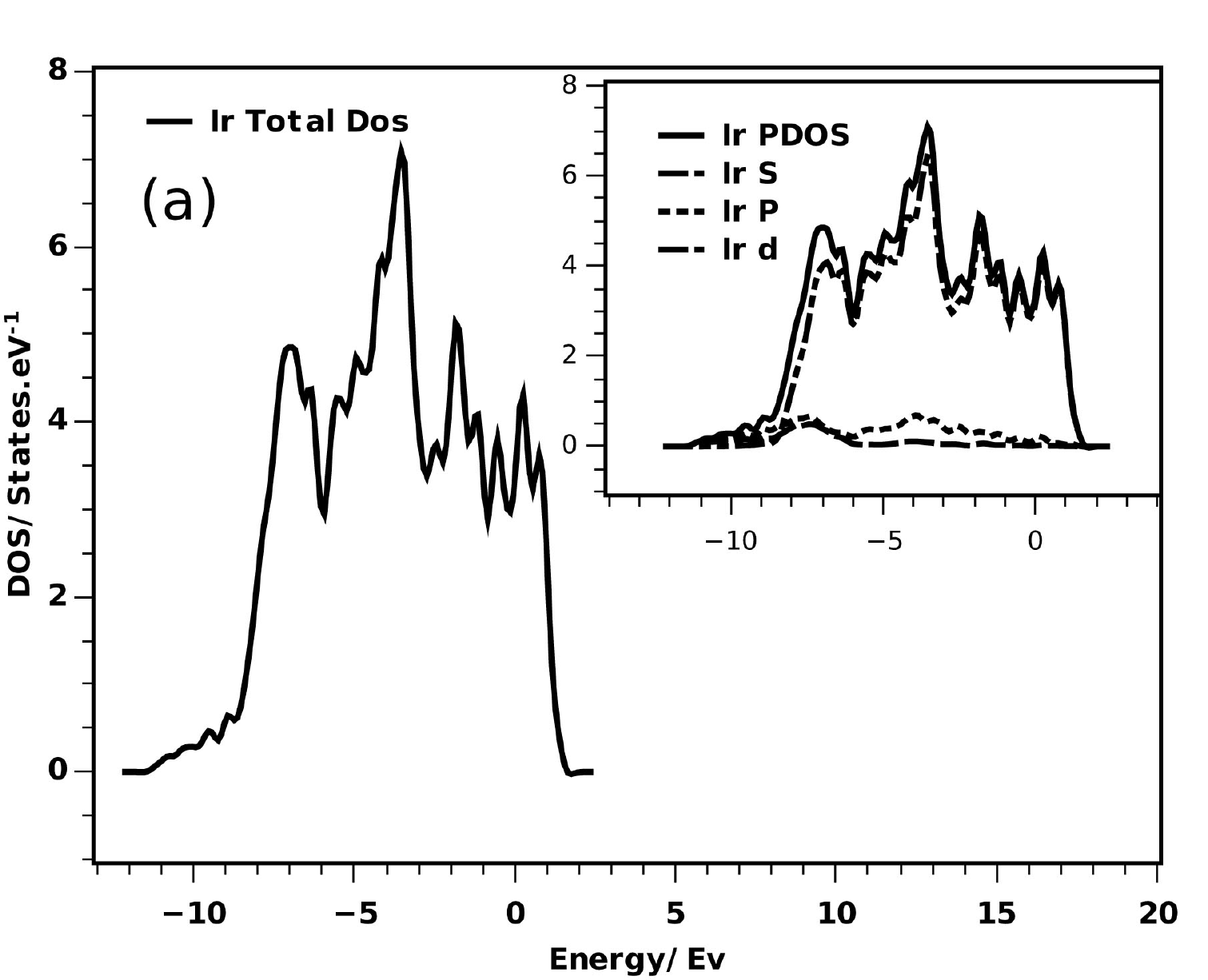}%
\hfill%
\includegraphics[width=0.46\textwidth]{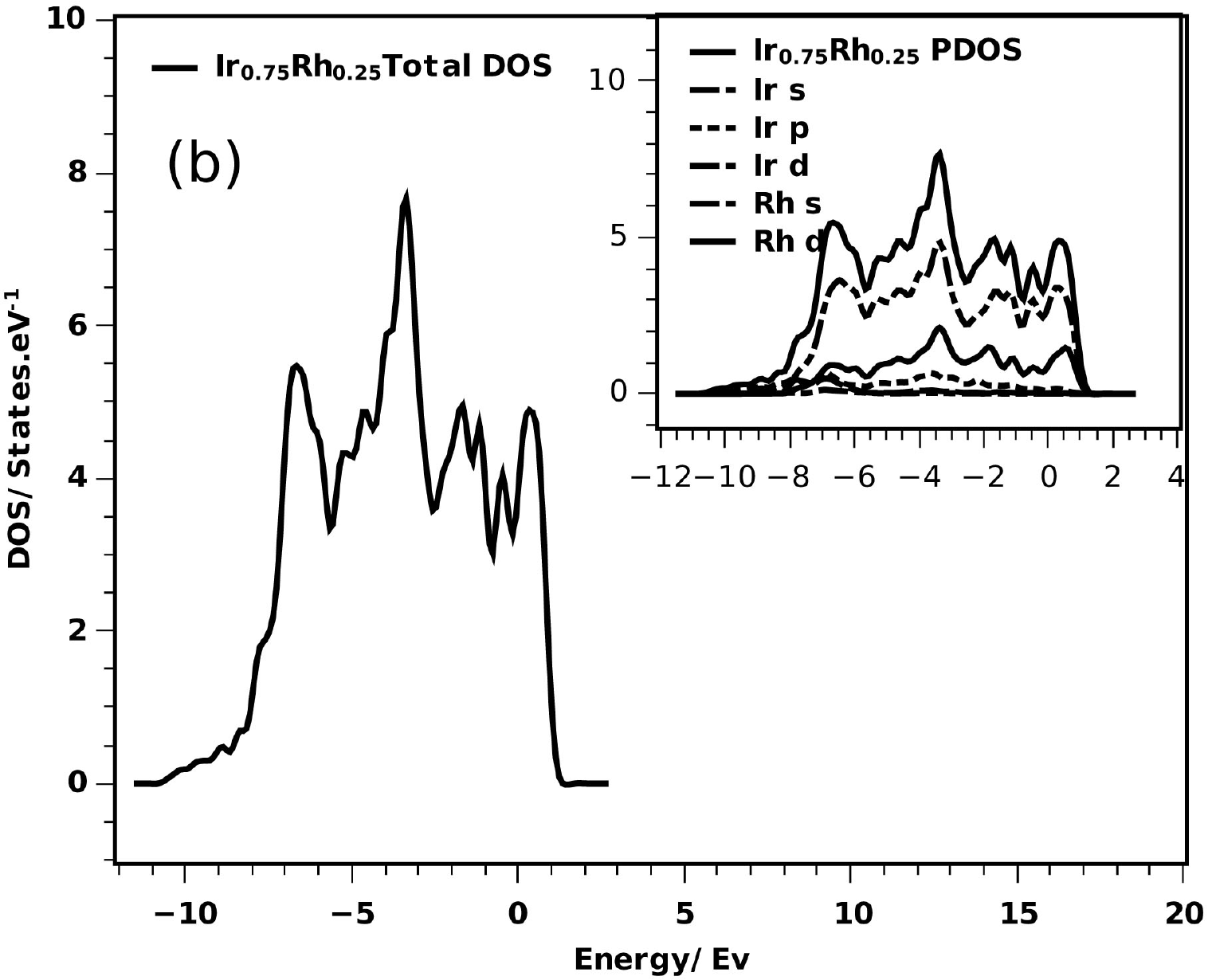}
\hspace{3mm}
\\
\includegraphics[width=0.48\textwidth]{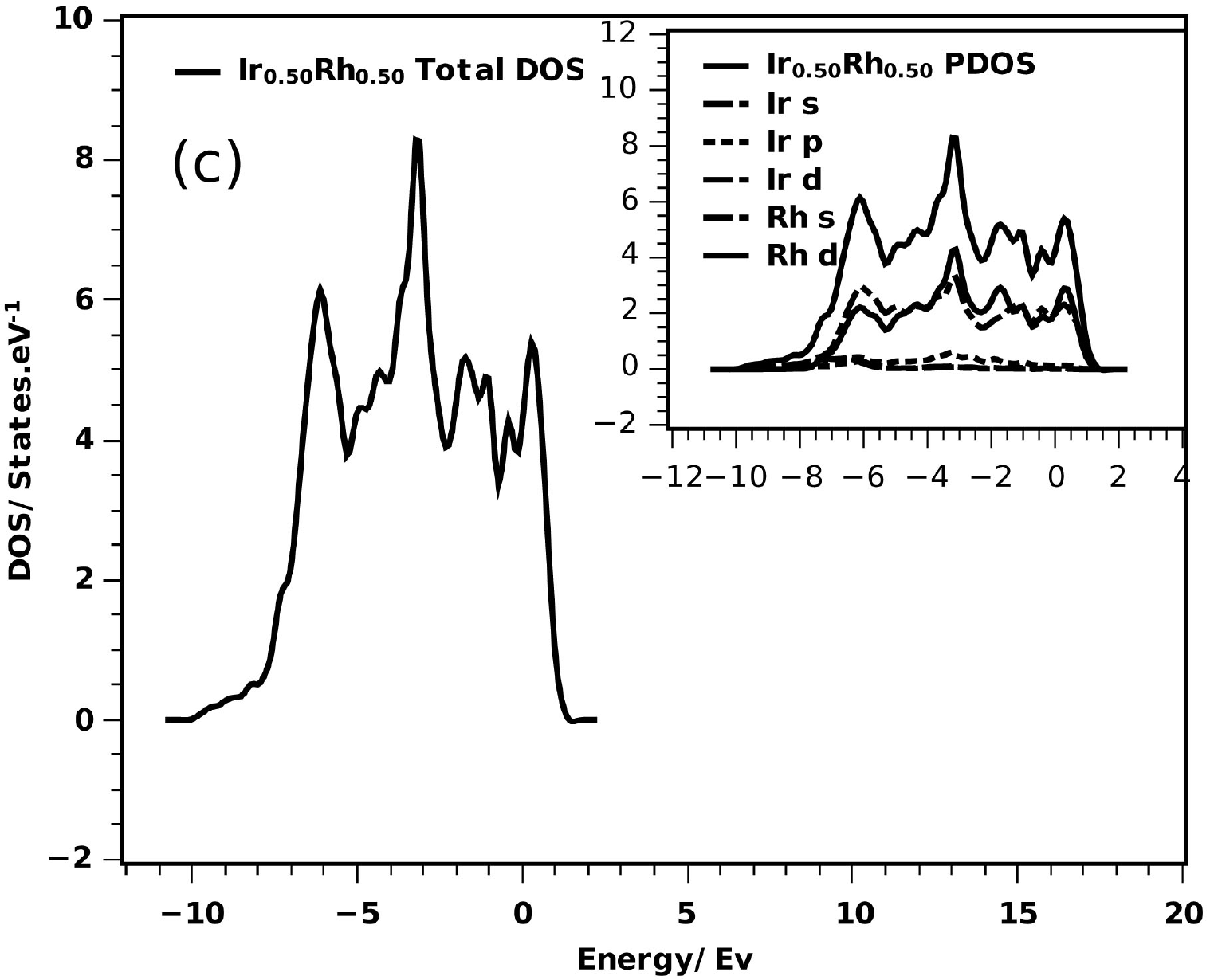}%
\hfill%
\includegraphics[width=0.48\textwidth]{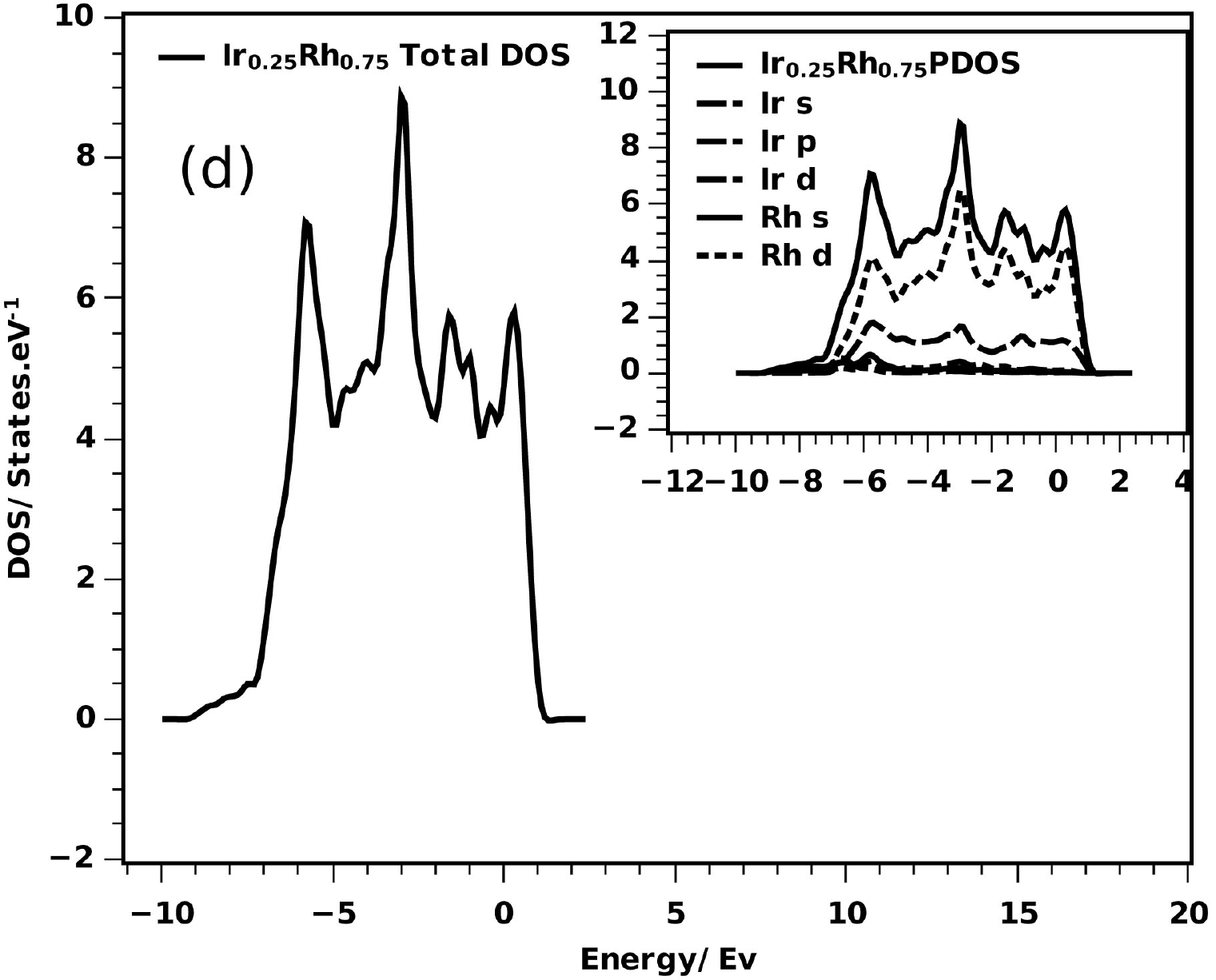}
\\
\centerline{\includegraphics[width=0.48\textwidth]{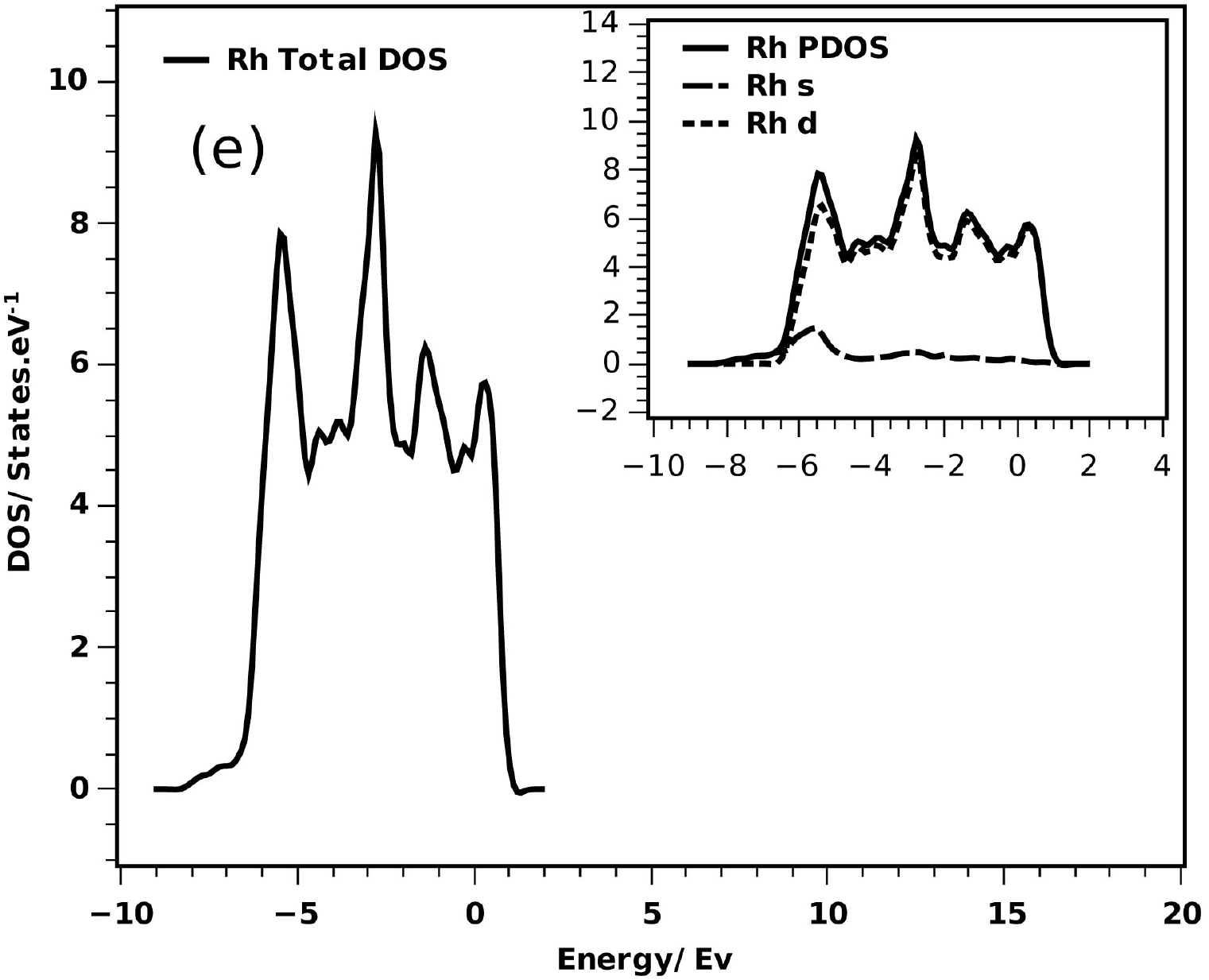}}
\caption{Density of states for Ir$_{1-x}$Rh$_{x}$ alloys: (a) at $x= 0.00$, (b) at $x= 0.25$, (c) at $x= 0.50$, (d) at $x= 0.75$, and (e) at $x= 1.00$ concentrations.}
\label{fig4}
\end{figure}

The density of states plays an essential role in studying many physical properties. Total energies of material can be calculated from the knowledge of the density of states. The number of electrons within the Fermi surface can be used to determine the nature of materials. The Fermi energy is determined by the Density of States (DOS) which provides information on the Fermi energy level.
To further study the nature of electronic band structure, we have also calculated the total and partial DOS for these alloys for different concentrations at an ambient pressure and presented them in figure~\ref{fig4}~(a)--(e). In figure~\ref{fig4}~(a), for pure Ir, the lowest lying bands were due to $s$, $p$ states and the higher energy states were mainly due to $d$ states, and the conductivity was dominated by $s$, $p$ and $d$ hybridization states. In case of Ir$_{0.75}$Rh$_{0.25}$ in figure~\ref{fig4}~(b), the bottom state of the valence band was dominated with $s$, $p$ states of Ir and Rh $s$ state. The upper valance band region was dominated with Ir $d$ state with some contribution of other states. In figure~\ref{fig4}~(c), with Ir$_{0.50}$Rh$_{0.50}$, the lowest part of the valance band was mostly contributed by $s$ state of both metals and Ir $p$ state but the higher energy level was dominated by a broad contribution of $d$-states of these metals with little part of other states. In figure~\ref{fig4}~(d), for Ir$_{0.25}$Rh$_{0.75}$, $s$, $p$ states of Ir and Rh $s$ state for this alloys dominated the bottom of the valence band. In the upper band, Rh $d$ state was much wider than the Ir $d$ states. In figure~\ref{fig4}~(e), with pure Rh, the bottom valence levels were mostly dominated by $s$ states but the higher valence  band was dominated with $d$ states.

From the analysis of the band structure of Ir$_{1-x}$Rh$_{x}$ alloy, it was observed that, at different concentrations of these metals,  the band structure  shows a  similar behavior  that  differs from each other mainly by the energy level of each band relative to the Fermi level. In pure Ir, the conductivity was mostly dominated by $s$, $p$ and $d$ hybridization but for pure Rh, the conductivity was prominent with $s$ and $d$ overlapping. The electronic energy density of states are compared with the previously reported result for cubic Ir and Rh \cite{46,47}.
When these metals were mixed to make alloys, strong hybridization between the Ir and Rh $d$ states with small contribution of $s$ states dominated the conductivity. Ir has a lower electronic DOS than Rh at the Fermi level and it increases with an increase of Rh concentration in the alloy. The material having a lower electronic DOS at the Fermi level often characterizes a more stable structure \cite{48,49}. Thus, Ir is a more stable structure than Rh and it decreases with Rh concentration. 	

\subsection{Thermal properties}

Thermal properties of metals are of interest below the melting temperature. We have used a quasi-harmonic approximation and phonon density of states to obtain thermal properties. Helmholtz free energy $\Delta F$, internal energy $\Delta E$, entropy $S$ and constant-volume specific heat $C_v$, at zero pressure are calculated.
 \begin{figure}[!h]
\includegraphics[width=0.48\textwidth]{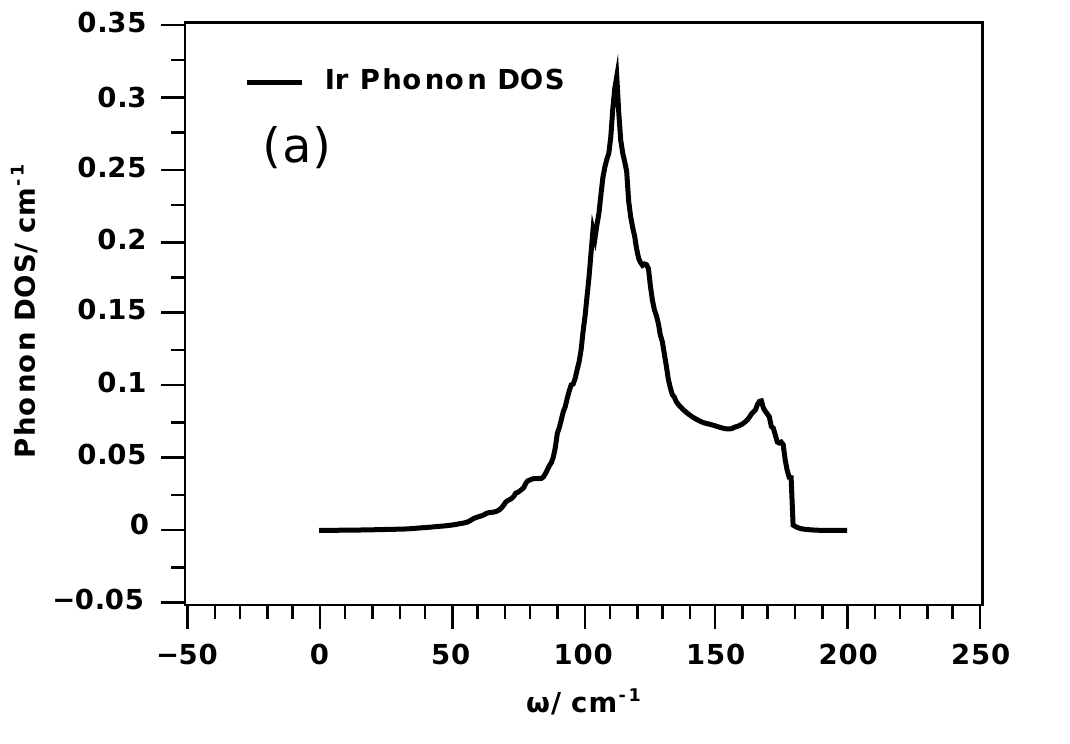}%
\hfill%
\includegraphics[width=0.48\textwidth]{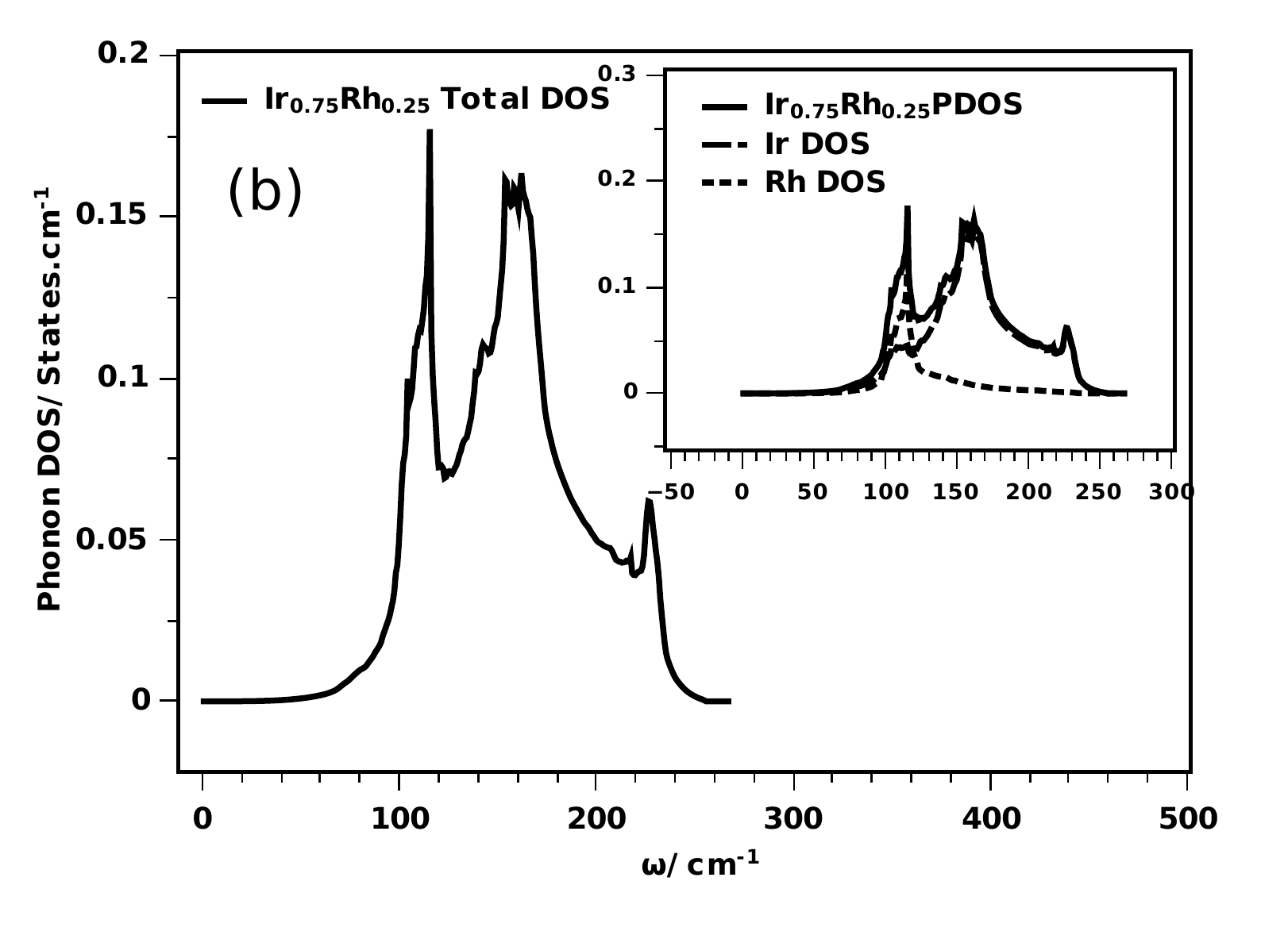}
\\
\includegraphics[width=0.48\textwidth]{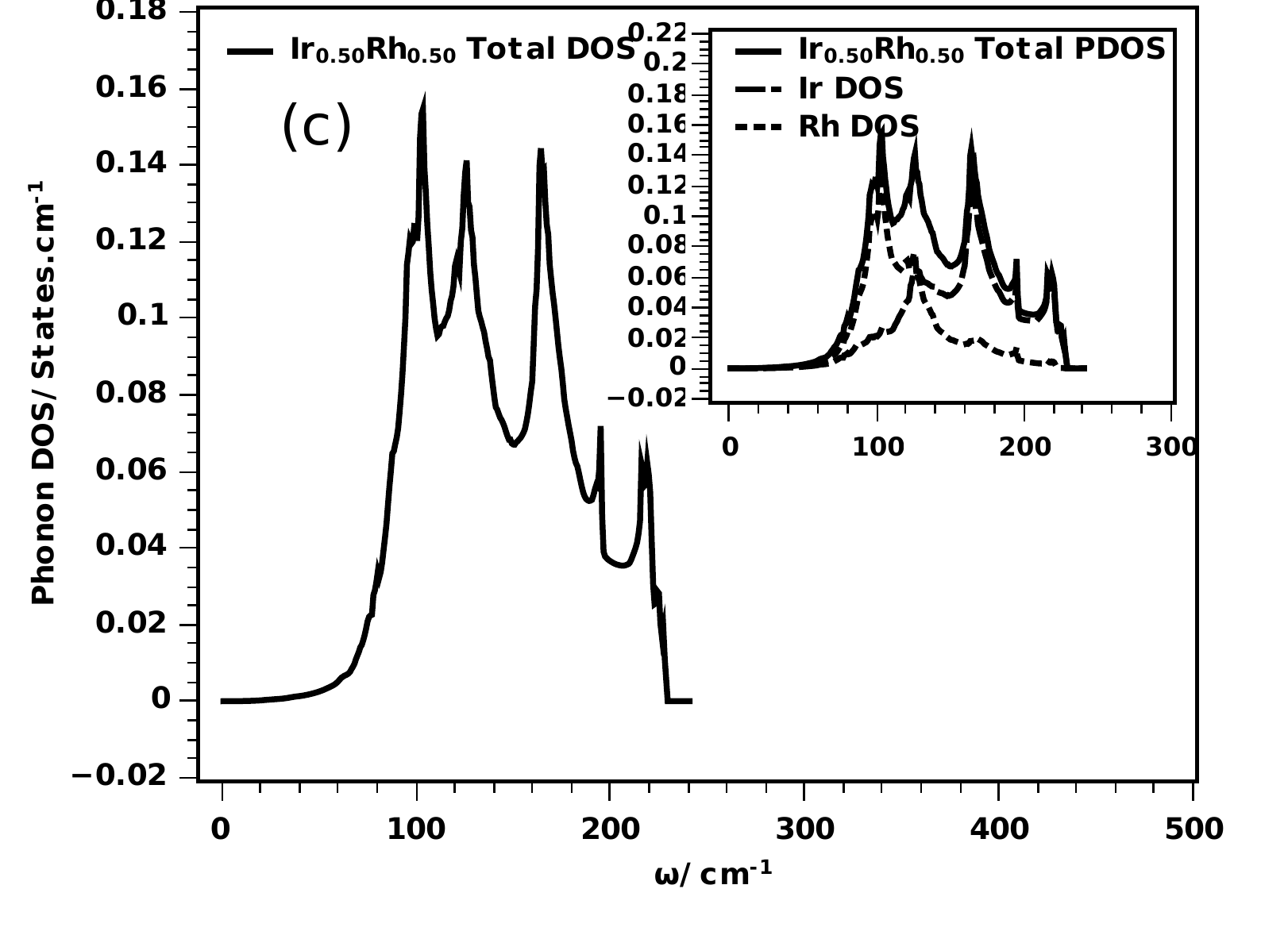}%
\hfill%
\includegraphics[width=0.48\textwidth]{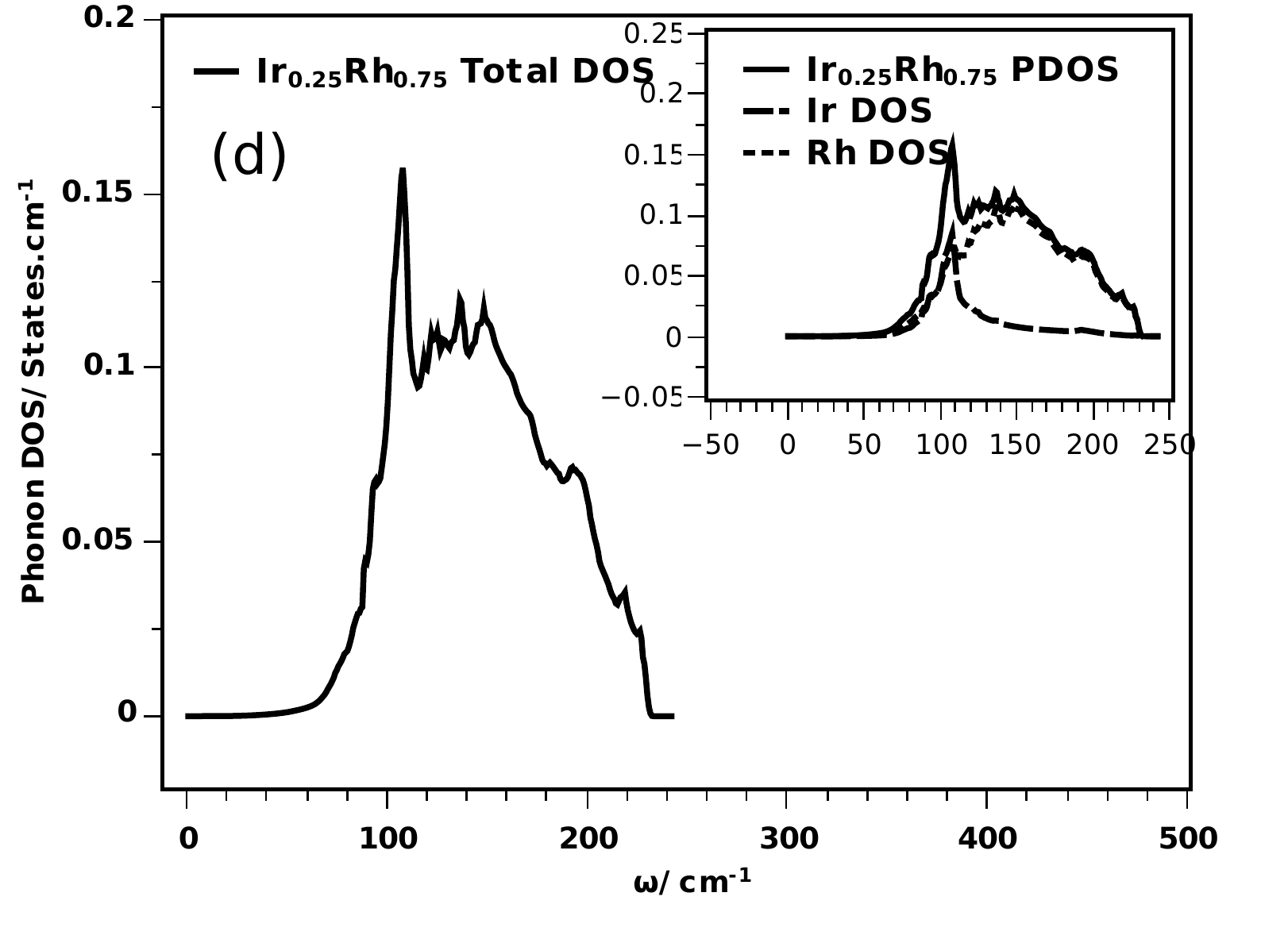}
\\
\centerline{\includegraphics[width=0.48\textwidth]{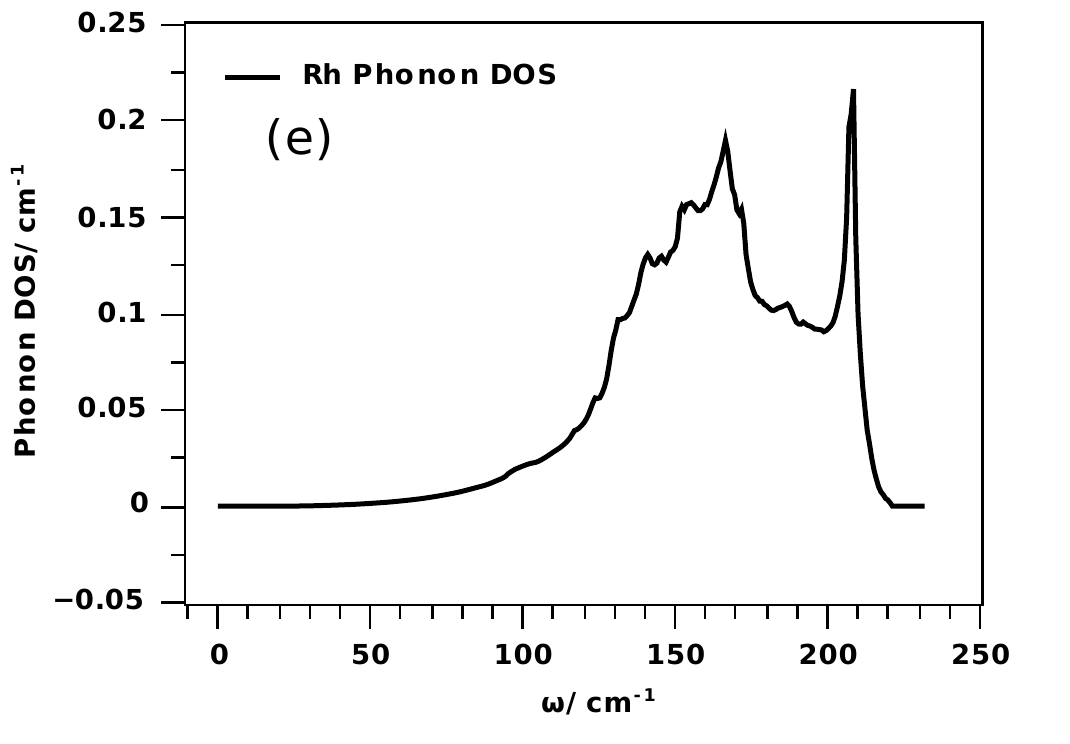}}
\caption{Phonon density of states for Ir$_{1-x}$Rh$_{x}$ alloys: (a) at $x= 0.00$, (b) at $x= 0.25$, (c) at $x= 0.50$, (d) at $x= 0.75$, and (e) at $x= 1.00$ concentrations.}
\label{fig5}
\end{figure}
Phonon DOS has a strong impact on thermodynamic properties. The electron excitation becomes easier with higher phonon DOS \cite{50}.
Figure~\ref{fig5} depicts the  phonon density of states of Ir$_{1-x}$Rh$_{x}$ alloys with different concentrations. The total phonon DOS for pure Ir was shown in figure~\ref{fig5}~(a) and has a maximum DOS at frequency 112.5~cm$^{-1}$. In figure~\ref{fig5}~(b), two higher peaks in phonon DOS occur, first peak is dominated by Rh at frequency of 115.5~cm$^{-1}$ and the second was due to Ir atoms at 162~cm$^{-1}$ frequency. In figure~\ref{fig5}~(c), three high peaks for phonon density of states occurs at a frequency of 103.5~cm$^{-1}$, 126~cm$^{-1}$, 164.25~cm$^{-1}$ which are due to movements of Rh, Ir and Rh and Ir atoms, respectively.  In figure~\ref{fig5}~(d), the contribution of total phonon DOS was dominated by Ir and Rh atoms at low frequency of 108~cm$^{-1}$. Conversely, phonon density of states is mainly dominated by Rh atoms after this frequency. In figure~\ref{fig5}~(e), for pure Rh, two higher peaks occur in phonon DOS, the first peak is dominated mainly at 155.25~cm$^{-1}$ frequency and the second peak with frequency 166.5~cm$^{-1}$.

\begin{figure}[!ht]
\includegraphics[width=0.48\textwidth]{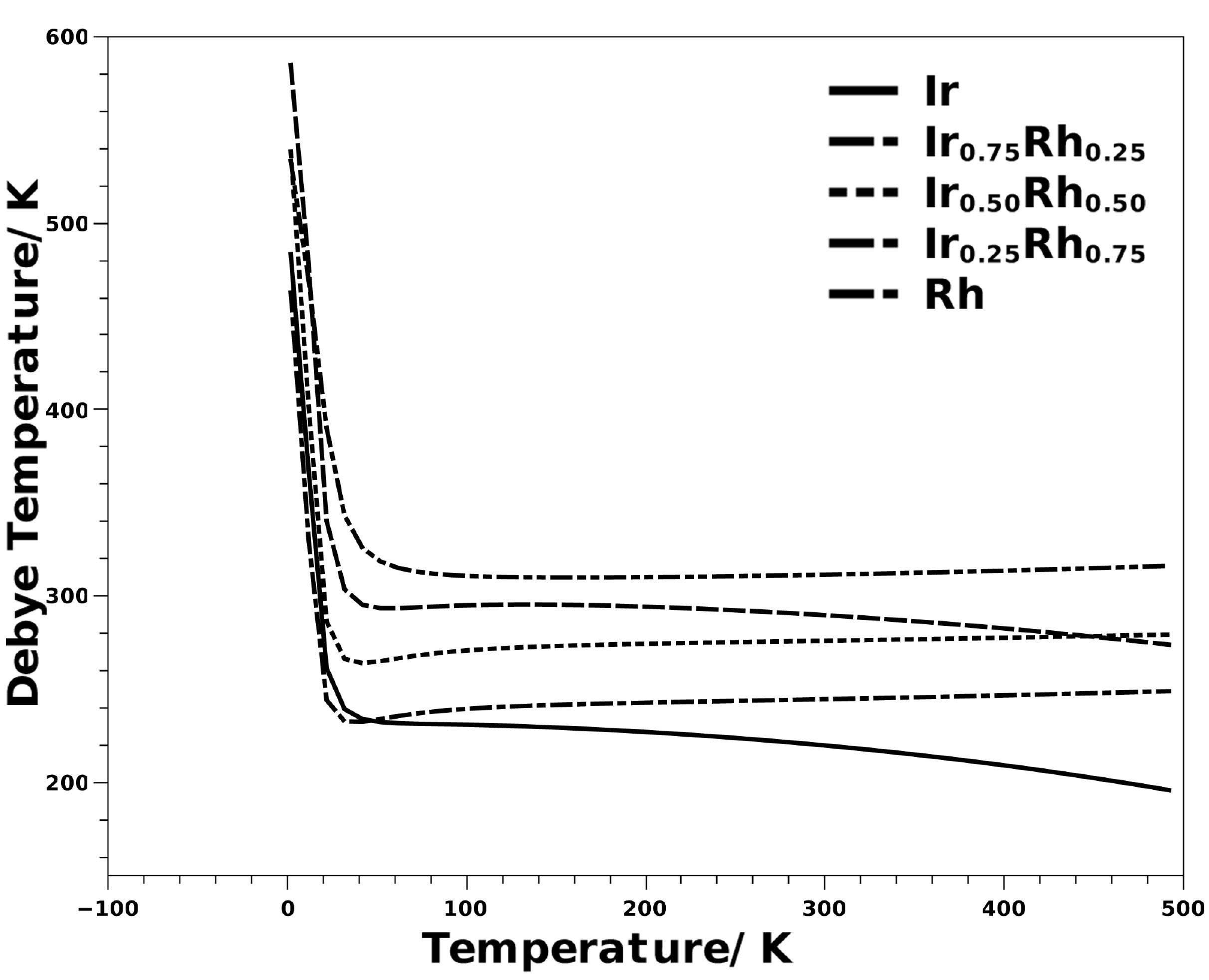}%
\hfill%
\includegraphics[width=0.48\textwidth]{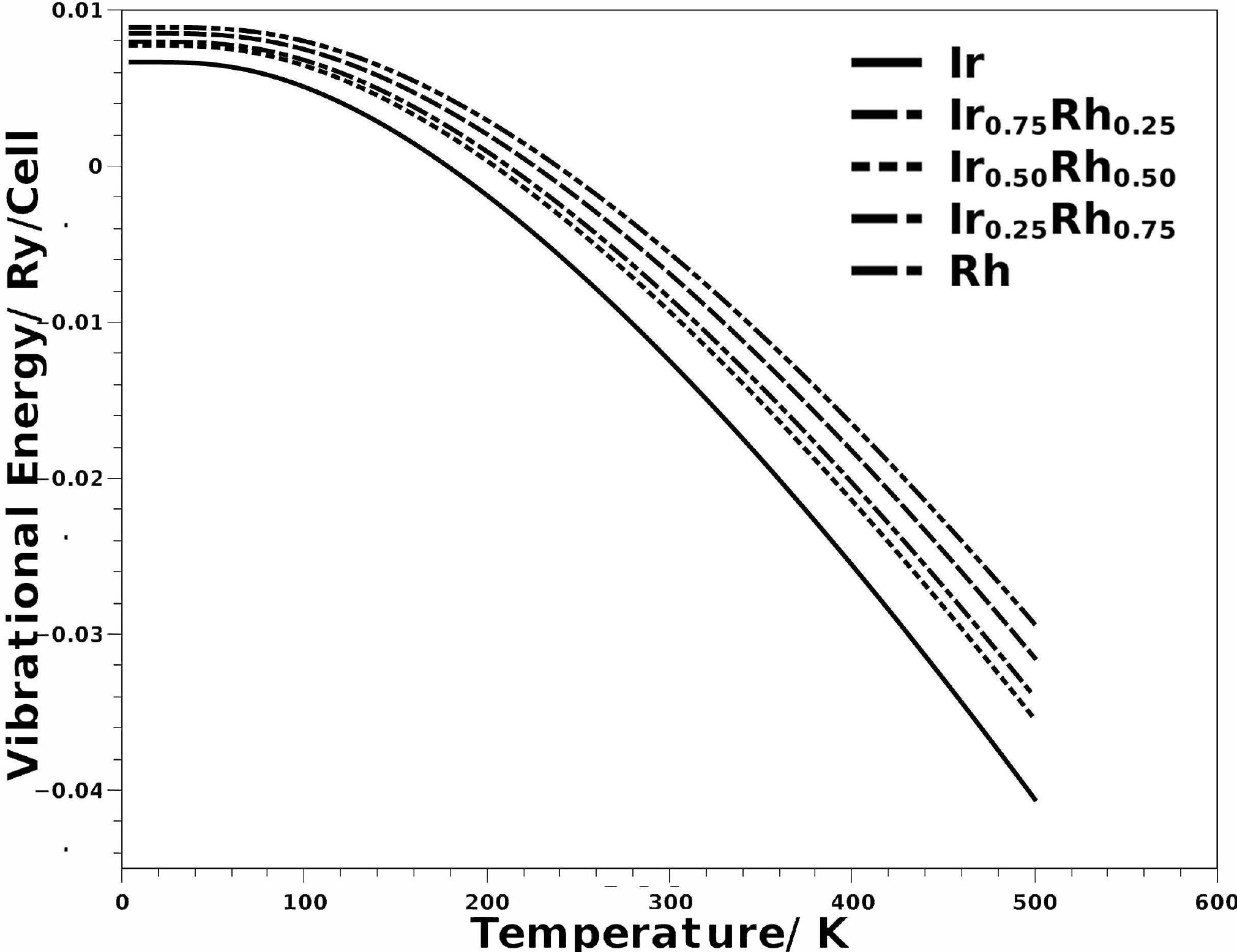}%
\\%
\parbox[t]{0.48\textwidth}{%
\caption{%
Debye temperature variation with temperature for  Ir$_{1-x}$Rh$_{x}$ alloys at $x= 0.00$, 0.25, 0.50, 0.75, 1.00.}
\label{fig6}%
}%
\hfill%
\parbox[t]{0.48\textwidth}{%
\caption{%
Vibration energy variation with temperature for Ir$_{1-x}$Rh$_{x}$ alloys at $x= 0.00$, 0.25, 0.50, 	0.75, and 1.00.}
\label{fig7}%
}%
\end{figure}
\begin{figure}[!h]
\includegraphics[width=0.48\textwidth]{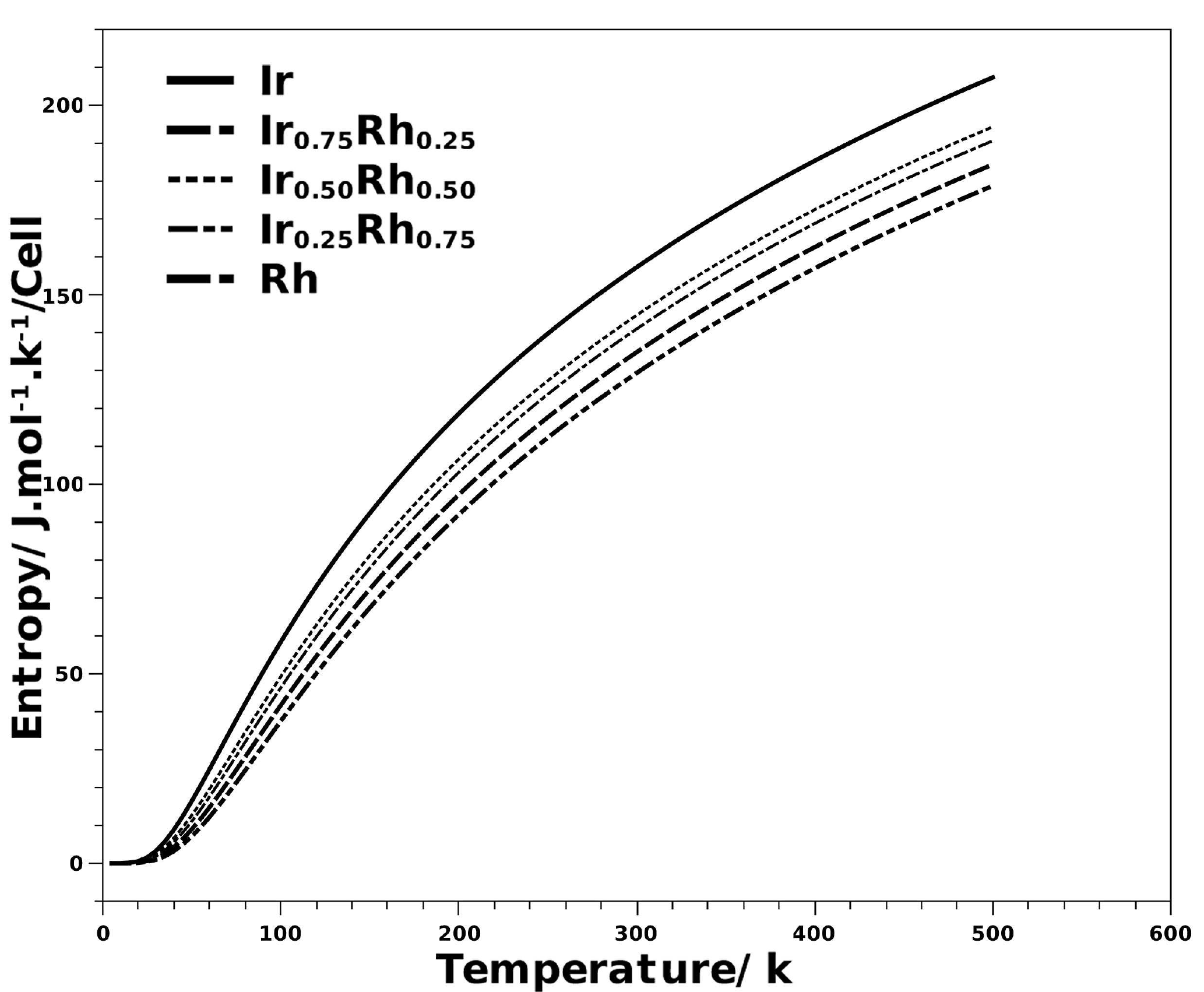}%
\hfill%
\includegraphics[width=0.48\textwidth]{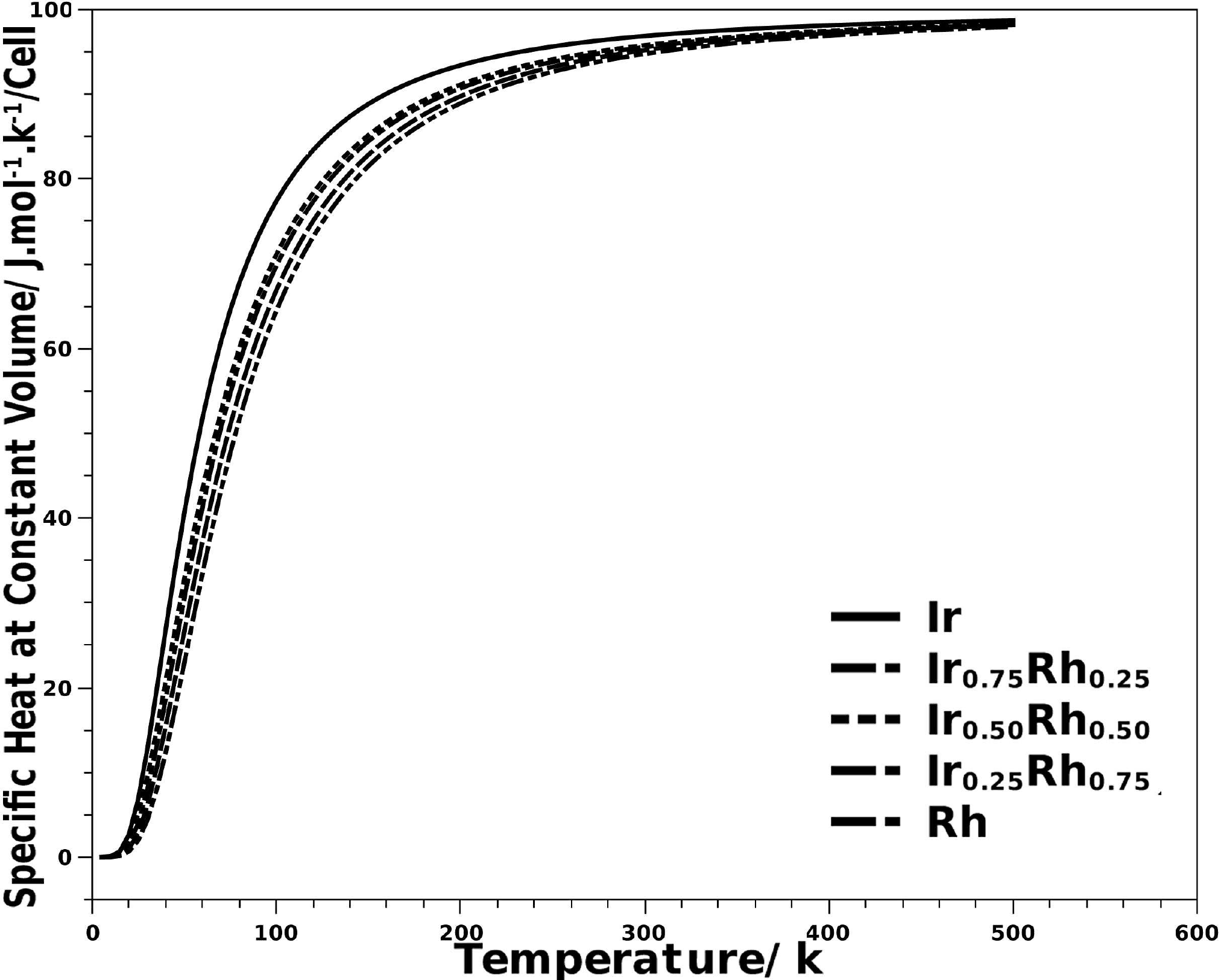}%
\\%
\parbox[t]{0.48\textwidth}{%
\caption{%
Entropy variation with temperature for Ir$_{1-x}$Rh$_{x}$alloys at $x= 0.00$, 0.25, 0.5, 0.75, and 	1.00.}
\label{fig8}%
}%
\hfill%
\parbox[t]{0.48\textwidth}{%
\caption{%
Specific heat at a constant volume variation with temperature for  Ir$_{1-x}$Rh$_{x}$ alloys at $x=0.00$, 0.25, 0.50, 0.75, and 1.00.}
\label{fig9}%
}%
\end{figure}
\begin{figure}[!h]
			\centerline{\includegraphics[width=0.48\textwidth]{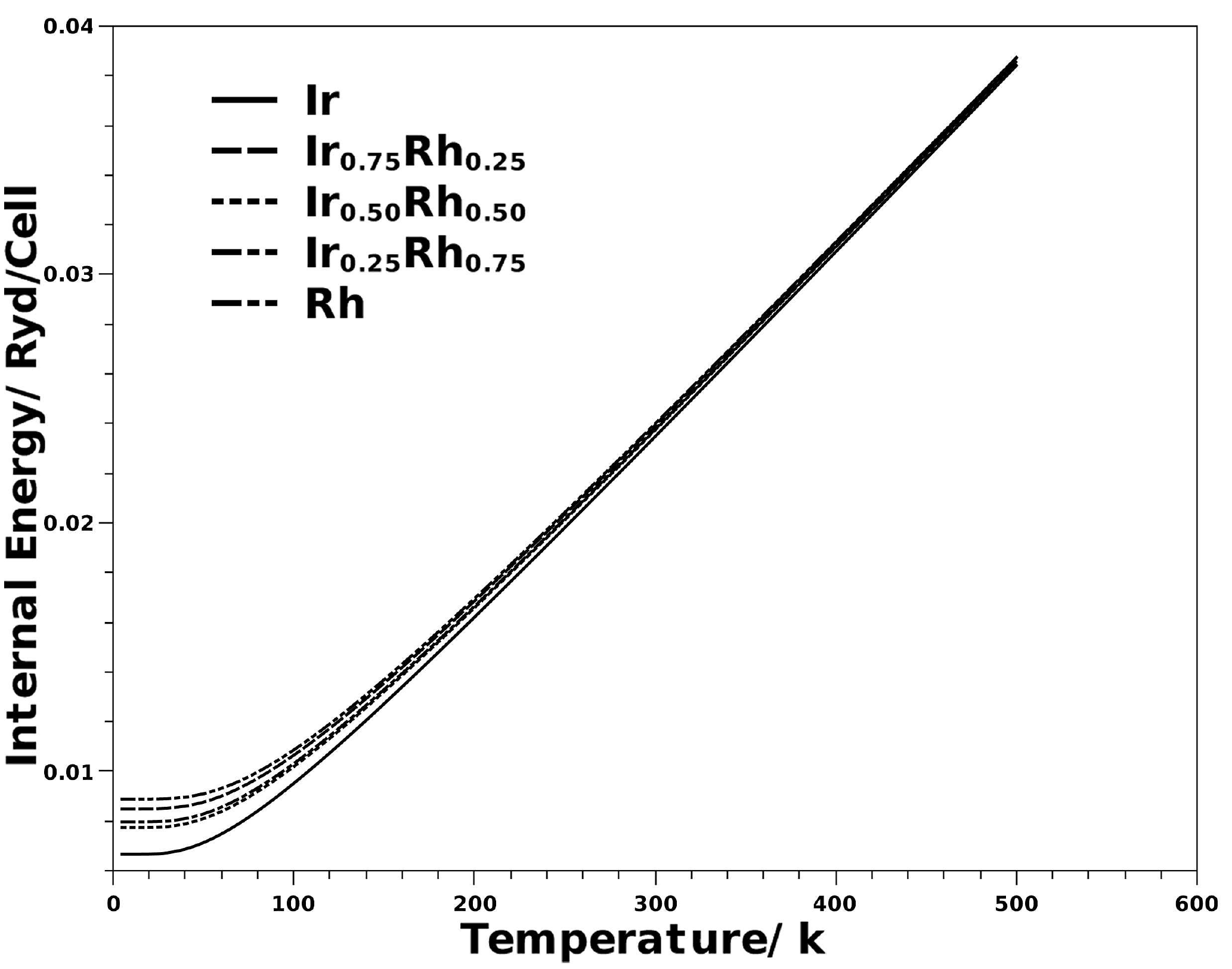}}
			\caption{Internal energy variation with temperature  for  Ir$_{1-x}$Rh$_{x}$ alloys at $x= 0.00$, 0.25, 0.50, 0.75, and 1.00.}
\label{fig10}
\end{figure}

To further study the thermal properties, the Debye temperatures, vibration energy, entropy, constant-volume specific heat and internal energy are plotted in figures~\ref{fig6}--\ref{fig10} for this alloy. Debye temperature determines the thermal characteristics of materials and it is closely related to many physical properties such as specific heat \cite{51}. The materials with high Debye temperature are associated with higher thermal conductivity. The knowledge of thermal conductivity and melting temperature is essential for developing and manufacturing electronic devices \cite{52,53}.  The Debye temperature plays an important role in the field of thermo-physical properties of materials. Debye temperature was maximum at 0--20~K, then it decreases rapidly and approaches a constant for higher temperature values. Entropy, constant-volume specific heat and internal energy values increase quickly at a lower temperature and becomes converged to constant values with high temperature as in figure~\ref{fig9}. Vibration energy remains constant for low temperature and decreases inversely as temperature increases. Entropy, constant-volume specific heat  decreases with Rh concentration but increases for Ir$_{0.50}$Rh$_{0.50}$, then continues the decreasing behavior for further Rh concentration. The increasing behavior may be due to the maximum lattice mismatch at 50 percent concentration and these quantities decrease with the further increase in Rh amount. It is seen that the internal energy and vibration energy increases slightly with Rh concentration but decreases for Ir$_{0.50}$Rh$_{0.50}$ concentration and then continues to progress. From figure~\ref{fig7}, it can also be seen that Rh has a greater vibration energy  compared to  Ir. However, entropy, constant-volume specific heat , internal energy, vibration energy and Debye temperature show a different behavior at 50 percent concentration which may be due to a greater lattice mismatch. High entropy possesses many attractive properties, such as high hardness \cite{54}, outstanding wear resistance \cite{55}, good fatigue resistance characteristics \cite{56}, excellent high-temperature strength \cite{57}, good thermal stability \cite{58} and, in general, good oxidation and corrosion resistance \cite{59}. These properties suggest a great potential in a wide variety of applications.

Our calculated results of thermal properties for  pure Ir and Rh are in good agreement with the previously published data \cite{60,61,62}. There were no experimental or theoretical results for their alloys.  Therefore,  the results reported in this work will be predictive for future. From the above investigations of thermodynamics properties, it can be demonstrated that Ir$_{1-x}$Rh$_{x}$ with $x= 0.5$ has a larger entropy and constant-volume specific heat due to a maximum lattice mismatch within the considered range of temperatures but the vibration energy and  the Debye temperature were minimum for $x= 0.5$.

\section{Conclusions}

The first principle method was employed  to analyze the structural, electronic, mechanical, and thermal properties of the Ir$_{1-x}$Rh$_{x}$ alloys with a four atom unit cell. The  pseudopotential scheme was used to study the alloys at different concentrations ($x = 0.00$, 0.25, 0.50, 0.75, 1.00).  The  following conclusions were drawn:
\begin{enumerate}
\item	 For the Ir$_{1-x}$Rh$_x$ alloys, by increasing Rh concentration, the lattice constant and the bulk modulus decrease. The variations in the calculated lattice constant and the bulk modulus are slightly different from Vegard's law.
\item	In electronic properties, the electronic bands overlap at the Fermi level, and the Fermi energy   decreases with an increase in Rh concentration but the band overlap increases. This is because  Rh has a larger electronic conductivity than Ir. In pure Ir, a narrow $d$ state was overlapped by a broad free-electron $s$ and $p$ but  pure Rh  has an overlap between $s$ and $d$ only which dominates the conductivity in these metals. When Ir and Rh were  mixed to form alloys, the main  hybridization was  between $d$ states  with some contribution from other states.
\item	Thermodynamic properties such as phonon density of states, the Helmholtz free energy, the phonon contribution to the internal energy, the constant-volume specific heat, and entropy were calculated using the quasi-harmonic approximation. The minimum value of phonon density of states is for the alloy having Ir$_{0.50}$Rh$_{0.50}$ concentration. Debye temperature was maximum at 0--20~K, then it decreases rapidly and approaches a constant for higher temperature values. Entropy, constant-volume specific heat and internal energy values increase quickly at a lower temperature and become converged to constant values with high temperature.  Entropy, constant-volume specific heat  decreases with Rh concentration but increases for Ir$_{0.50}$Rh$_{0.50}$, then continues a decreasing behavior for further Rh concentration. Vibration energy remains constant for low temperature and decreases inversely as temperature increases. It is seen that the internal energy and the vibration energy increase slightly with Rh concentration but decrease for Ir$_{0.50}$Rh$_{0.50}$ concentration and then continue to progress.
\end{enumerate}



\newpage
\ukrainianpart

\title{Ab initio дослідження  структурних електронних і теплових властивостей сплавів  Ir$_{1-x}$Rh$_{x}$  alloys}%

\author{Ш. Ахмед , М. Зафар, М. Шакіл, М.А. Чоуджарі} %

\address{
Лабораторія комп'ютерного моделювання, фізичний факультет, Ісламський університет м. Бахавалпур, м. Бахавалпур 63100, Пакистан
}%

\makeukrtitle
\begin{abstract}
Структурні, електронні, механічні і теплові властивості сплавів  Ir$_{1-x}$Rh$_{x}$ систематично досліджено з використанням
ab initio теорії функціоналду густини при різних концентраціях ($x = 0.00$, 0.25, 0.50, 0.75, 1.00).
Було використано  метод спеціальної квазівипадкової структури для моделювання  сплавів, що мають структуру FCC з чотирма атомами
на одиничну комірку. Були обчислені властивості основного стану, такі як постійна ґратки та об'ємний модуль пружності, для того, щоб
знайти рівноважні положення атомів для стійких сплавів.
Обчислені властивості основного стану добре узгоджуються  з експериментальними та іншими раніше отриманими теоретичними даними.
З метою вивчення електронних властивостей цих сплавів при різних концентраціях обчислено електронну зонну структуру та густину станів.
Електронні властивості обгрунтовують металічну поведінку сплавів. Для обчислення теплових характеристик була використана
першопринципна теорія  збурень функціоналу густини, імплементована у квазігармонічне наближення.  Нами обчислено такі теплові
характеристики, як температура Дебая, енергія коливань, ентропія, питома теплоємність при постійному об'ємі та внутрішня енергія.
Було використано ab initio метод лінійного відгуку для обчислення густини станів фононів.
\keywords електронні, стактурні і теплові властивості металів платинової групи
\end{abstract}


\begin{thebibliography}{99}

\bibitem{1}{Yuan Y., Yan N., Dyson P.J., ACS Catal., 2012, \textbf{2}, 1057; \doi{10.1021/cs300142u}.}

\bibitem{2}{Savitskii E.M., Handbook of Precious Metals,  Hemisphere, New York, 1989.}

\bibitem{3}{Cawkwell M.J., Nguyen-Manh D., Woodward C., Pettifor D.G., Vitek V., Science, 2005, \textbf{309}, 1059; \doi{10.1126/science.1114704}.}

\bibitem{4}{Stevanovi\'c V., Sljivancanin Z., Baldereschi A., Phys. Rev. Lett., 2007, \textbf{99}, 165501; \doi{10.1103/PhysRevLett.99.165501}.}

\bibitem{5}{Mumataz K., Echigoya J., Hirai T., Shindo  Y., Mat. Sci. Eng. A, 1993, \textbf{167}, 187; \doi{10.1016/0921-5093(93)90353-G}.}

\bibitem{6}{Kovacs G.T.A., Storment C.W.,  Kounaves  S.P., Sens. Actuators B, 1995, \textbf{23}, 41; \doi{10.1016/0925-4005(94)01523-K}.}

\bibitem{7}{Hamilton J.C., Yang N.Y.C., Clift W.M., Boehme D.R., McCarty K.F., Franklin J.E., Metall. Trans. A, 1992, \textbf{23}, 851; \doi{10.1007/BF02675562}.}

\bibitem{8}{Yet L., Chem. Rev., 2000, \textbf{100}, 2963; \doi{10.1021/cr990407q}.}

\bibitem{9}{Progress in Inorganic Chemistry, vol. 28, Lippard S.J. (Ed.), Wiley, New York, 1981.}

\bibitem{10}{Rossen K.,  Angew. Chem. Int. Ed., 2001, \textbf{40},  24; \doi{10.1002/1521-3773(20011217)40:24\%3C4611::AID-ANIE4611\%3E3.3.CO;2-W}.}

\bibitem{11}{Ungv\'ary F., Coord. Chem. Rev., 2002, \textbf{228}, 61; \doi{10.1016/S0010-8545(02)00051-6}.}

\bibitem{12}{Yamabe Y., Koizumi Y., Murakami H., Maruko T., Harada H., Scr. Mater., 1996, \textbf{35}, 211; \doi{10.1016/1359-6462(96)00109-1}.}

\bibitem{13}{Yamabe Y., Koizumi Y., Murakami H., Maruko T., Harada H., Scr. Mater.,  1997, \textbf{36}, 393; \doi{10.1016/S1359-6462(96)00408-3}.}

\bibitem{14}{Miura S., Honmab K., Teradac Y.,  Sanchezd J.M., Mohria T., Intermetallics, 2000, \textbf{8}, 785; \doi{10.1016/S0966-9795(00)00012-1}.}

\bibitem{15}{Iotova D., Kioussis N., Lim S.P., Phys. Rev. B, 1996, \textbf{54}, 14413; \doi{10.1103/PhysRevB.54.14413}.}

\bibitem{16}{Yamabe Y., Koizumi Y., Murakami H., Maruko T., Harada H., Scr. Mater., 1996, \textbf{35}, 211; \doi{10.1016/1359-6462(96)00109-1}.}

\bibitem{17}{Yamabe Mitarai Y., Ro Y., Maruko T., Harada H., Metall. Mater. Trans. A, 1998, \textbf{29}, 537; \doi{10.1007/s11661-998-0135-9}.}

\bibitem{18}{Yamabe-Mitarai Y., Harada H., Gu Y., Huang C., Metall. Mater. Trans. A, 2005, \textbf{36}, 547; \doi{10.1007/s11661-005-0169-1}.}

\bibitem{19}{Mansour A.N., Dmitrienko A., Soldatov A.V., Phys. Rev. B, 1997, \textbf{55}, 15531; \doi{10.1103/PhysRevB.55.15531}.}

\bibitem{20}{Marzari N., MRS Bull., 2006, \textbf{31}, 681; \doi{10.1557/mrs2006.177}.}

\bibitem{21}{Okoye C.M.I., J. Phys.: Condens. Matter, 2003, \textbf{15}, 5945; \doi{10.1088/0953-8984/15/35/304}.}

\bibitem{22}{Bernick R.L., Kleinman L.,  Solid State Commun., 1970, \textbf{8}, 569; \doi{10.1016/0038-1098(70)90305-4}.}

\bibitem{23}{Parr R.G., Yang W., Density-Functional Theory of Atoms and Molecules, Oxford University Press, Oxford, 1989.}

\bibitem{24}{Dreizler R.M., Gross E.K.U., Density functional theory, Springer, Berlin, 1990.}

\bibitem{25}{Ivanov A.S.,  Katsnelson M.I.,  Mikhin A.G.,  Osetskii Yu.N., Rumyantsev A.Yu., Trefilov~A.V.,  Shamanaev~Yu.F.,  Yakovenkova~L.I., Philos. Mag.~B, 1994, \textbf{69}, 1183; \doi{10.1080/01418639408240188}.}

\bibitem{26}{Baroni  S., Dal Corso A., Gironcoli S., Giannozzi P., Rev. Mod. Phys., 2001, \textbf{73}, 515; \doi{10.1103/RevModPhys.73.515}.}

\bibitem{27}{Ceperley  D.M., Alder B.J., Phys. Rev. Lett., 1980, \textbf{45}, 566; \doi{10.1103/PhysRevLett.45.566}.}

\bibitem{28}{Perdew J.P., Zunger A., Phys. Rev. B, 1981, \textbf{23}, 5048; \doi{10.1103/PhysRevB.23.5048}.}

\bibitem{29}{Vanderbilt D., Phys. Rev. B, 1985, \textbf{32},  8412; \doi{10.1103/PhysRevB.32.8412}.}

\bibitem{30}{Perdew J.P., Burke K., Ernzerhof M., Phys. Rev. Lett., 1996, \textbf{77}, 3865; \doi{10.1103/PhysRevLett.77.3865}.}

\bibitem{31}{Monkhorst H.J., Pack J.D., Phys. Rev. B, 1976, \textbf{13}, 5188; \doi{10.1103/PhysRevB.13.5188}.}

\bibitem{32}{Zunger A., Wei S.H., Ferreira L.G., Bernard J.E., Phys. Rev. Lett., 1990, \textbf{65}, 353; \doi{10.1103/PhysRevLett.65.353}.}

\bibitem{33}{Murnaghan F.D., Proc. Natl. Acad. Sci. USA, 1944, \textbf{30}, 244; \doi{10.1073/pnas.30.9.244}.}

\bibitem{34}{Mehl M.J., Papaconstantopoulos D.A., Phys. Rev. B, 1996, \textbf{54}, 4519; \doi{10.1103/PhysRevB.54.4519}.}

\bibitem{35}{Gong H.R., Scr. Mater., 2008, \textbf{59}, 1197; \doi{10.1016/j.scriptamat.2008.08.009}.}

\bibitem{36}{Singh H.P., Acta Crystallogr. A, 1968, \textbf{24}, 469; \doi{10.1107/S056773946800094X}.}

\bibitem{37}{G\'omez J., Ding Y., Koitz R., Seitsonen A.P., Iann M., Theor. Chem. Acc.,  2013, \textbf{132}, 1350; \doi{10.1007/s00214-013-1350-z}.}

\bibitem{38}{Simmons G., Wang H., Single Crystal Elastic Constants and Calculated Aggregate Properties, 2nd Edn., MIT Press, Cambridge, 1971.}

\bibitem{39}{Walker E., Ashkenazi J., Dacorogna M., Phys. Rev. B, 1981, \textbf{24}, 2254; \doi{10.1103/PhysRevB.24.2254}.}

\bibitem{40}{Niranjan M.K., Intermetallics, 2012, \textbf{26}, 150; \doi{10.1016/j.intermet.2012.03.049}.}

\bibitem{41}{Sa'adi H., Hamad B, J. Phys. Chem. Solids, 2008, \textbf{69}, 2457; \doi{10.1016/j.jpcs.2008.04.038}.}

\bibitem{42}{Hollister R.G.,  Darling A.S., Platin. Met. Rev., 1967, \textbf{11}, No.~3, 94.}

\bibitem{43}{Ahuja B.L., Sharma V., Rathor A., Jani A.R., Sharma B.K., Nucl. Instrum. Methods Phys. Res., Sect. B, 2007, \textbf{262}, 391; \doi{10.1016/j.nimb.2007.05.029}.}

\bibitem{44}{Noffke J., Fritsche L.,  J. Phys. F: Met. Phys., 1982, \textbf{12}, 921; \doi{10.1088/0305-4608/12/5/011}}.

\bibitem{45}{Powell R.W.,   Tye R.P.,  Woodman M.J., Platin. Met. Rev., 1962, \textbf{6}, No.~4, 138.}

\bibitem{46}{Bannikov V.V., Shein I.R., Ivanovskii A.L., Solid State Commun., 2009, \textbf{149}, 1807; \doi{10.1016/j.ssc.2009.07.015}.}

\bibitem{47}{Sudhapriyanga G., Asvinimeenaatcia A.T., Rajeswarapalanichamy  R.,  Iyakutti  K., Acta Phys. Pol. A, 2014, \textbf{125}, 29; \doi{10.12693/APhysPolA.125.29}.}

\bibitem{48}{Xu J.H., Freeman A., Phys. Rev. B, 1990, \textbf{41}, 12553; \doi{10.1103/PhysRevB.41.12553}.}

\bibitem{49}{Xu J.H., Oguchi T., Freeman A.J., Phys. Rev. B, 1987, \textbf{35}, 6940; \doi{10.1103/PhysRevB.35.6940}.}

\bibitem{50}{Liang C.P., Gong H.R., Intermetallics, 2013, \textbf{32}, 429; \doi{10.1016/j.intermet.2012.09.014}.}

\bibitem{51}{Boucetta S.,  Zegrar F., Journal of Magnesium and Alloys, 2013, \textbf{1}, 128; \doi{10.1016/j.jma.2013.05.001}.}

\bibitem{52}{Newnham R.E., Properties of Materials:
Anisotropy, Symmetry, Structure, Oxford University Press, New York, 2005.}

\bibitem{53}{Mattesini M., Magnuson M., Tasn\'{a}di F., H\"{o}glund C., Abrikosov I.A., Hultman L., Phy. Rev. B, 2009, \textbf{79}, 125122; \doi{10.1103/PhysRevB.79.125122}.}

\bibitem{54}{Chen S.T., Tang W.Y., Kuo Y.F., Chen S.Y., Tsau C.H., Shun~T.T., Yeh~J.W., Mater. Sci. Eng. A, 2010, \textbf{527}, 5818; \doi{10.1016/j.msea.2010.05.052}.}

\bibitem{55}{Chuang M.H., Tsai M.H., Wang W.R., Lin S.J., Yeh J.W., Acta Mater., 2011, \textbf{59}, 6308; \doi{10.1016/j.actamat.2011.06.041}.}

\bibitem{56}{Hemphill M.A., Yuan T., Wang G.Y., Yeh J.W., Tsai C.W., Chuang A., Liaw P.K., Acta Mater., 2012, \textbf{60}, 5723; \doi{10.1016/j.actamat.2012.06.046}.}

\bibitem{57}{Hsu C.Y., Juan C.C., Wang W.R., Sheu T.S., Yeh J.W., Chen S.K., Mater. Sci. Eng. A, 2011, \textbf{528}, 3581; \doi{10.1016/j.msea.2011.01.072}.}

\bibitem{58}{Tsai M.H., Wang C.W., Tsai C.W., Shen W.J., Yeh J.W., Gan, J.Y., Wu W.W., J. Electrochem. Soc., 2011, \textbf{158}, H1161; \doi{10.1149/2.056111jes}.}

\bibitem{59}{Chou Y.L., Wang Y.C., Yeh J.W., Shih H.C., Corros. Sci., 2010, \textbf{52}, 3481; \doi{10.1016/j.corsci.2010.06.025}.}

\bibitem{60}{The American Institute of Physics Handbook, Gray D.E. (Ed.), McGraw-Hill, New York, 1972.}

\bibitem{61}{Arblaster J.W.,  Platin. Met. Rev., 1996, \textbf{40}, 62.}

\bibitem{62}{Konti A., Varshni Y.P., Can. J. Phys., 1969, \textbf{47}, 2021; \doi{10.1139/p69-255}.}

\end{thebibliography}
\end{document}